\documentclass{article}
\usepackage[utf8]{inputenc}
\usepackage{amsmath}
\usepackage{amsfonts}
\usepackage{amssymb}

\usepackage[a4paper,includeheadfoot,margin=2.9cm,top=2.44cm,bottom=2.44cm]{geometry}

\usepackage{bbold}
\usepackage[parfill]{parskip}
\usepackage[frozencache]{minted}
\usepackage{etoolbox,xpatch}
\usepackage{newunicodechar}
\newunicodechar{δ}{\ensuremath{\delta}}
\newunicodechar{Γ}{\ensuremath{\Gamma}}
\newunicodechar{ε}{\ensuremath{\epsilon}}
\setmintedinline{breaklines}

\newcommand{\imgPlot}[1]{\includegraphics[scale=0.7]{Plots/Plot#1.pdf}}

\makeatletter
\AtBeginEnvironment{minted}{\dontdofcolorbox}
\def\dontdofcolorbox{\renewcommand\fcolorbox[4][]{##4}}
\xpatchcmd{\inputminted}{\minted@fvset}{\minted@fvset\dontdofcolorbox}{}{}
\xpatchcmd{\mintinline}{\minted@fvset}{\minted@fvset\dontdofcolorbox}{}{} % see https://tex.stackexchange.com/a/401250/
\makeatother

\usepackage{tabularx}

\usepackage{graphicx}
\usepackage[hyperindex,breaklinks,hidelinks]{hyperref}
\usepackage{xcolor}

\definecolor{airforceblue}{rgb}{0.26, 0.44, 0.56}

\usepackage{mathtools}

\usepackage{mathtools}
\DeclarePairedDelimiter{\floor}{\lfloor}{\rfloor}

\usepackage{stackengine}
\stackMath
\newcommand\tsup[2][2]{%
 \def\useanchorwidth{T}%
  \ifnum#1>1%
    \stackon[-.5pt]{\tsup[\numexpr#1-1\relax]{#2}}{\scriptscriptstyle\sim}%
  \else%
    \stackon[.5pt]{#2}{\scriptscriptstyle\sim}%
  \fi%
}

\usepackage{multirow}
\usepackage{ltablex}

\newcommand\xrowht[2][0]{\addstackgap[.5\dimexpr#2\relax]{\vphantom{#1}}}

\usepackage{authblk}
\usepackage{cite}

\usepackage[htt]{hyphenat}

\usepackage{fancyhdr}
\fancypagestyle{plain}{%
	\fancyhead[R]{~\\TCDMATH 20-07}
	
}

\title{DiffExp, a Mathematica package for computing Feynman integrals in terms of one-dimensional series expansions
}

\author{Martijn Hidding \\
{\it School of Mathematics, Trinity College Dublin, Dublin 2, Ireland}\\

\vspace*{3.8mm} {E-mail:
hiddingm@tcd.ie
}
}

\date{}

\begin{document}

\maketitle

\begin{abstract}
DiffExp is a Mathematica package for integrating families of Feynman integrals order-by-order in the dimensional regulator from their systems of differential equations, in terms of one-dimensional series expansions along lines in phase-space, which are truncated at a given order in the line parameter. DiffExp is based on the series expansion strategies that were explored in recent literature for the computation of families of Feynman integrals relevant for Higgs plus jet production with full heavy quark mass dependence at next-to-leading order. The main contribution of this paper, and its associated package, is to provide a public implementation of these series expansion methods, which works for any family of integrals for which the user provides a set of differential equations and boundary conditions (and for which the program is not computationally constrained.) The main functions of the DiffExp package are discussed, and its use is illustrated by applying it to the three loop equal-mass and unequal-mass banana graph families.
\end{abstract}

\tableofcontents

\section{Introduction}

In this paper we present the Mathematica package DiffExp for the integration of Feynman integrals in terms of one-dimensional power series expansions from their systems of differential equations, which can obtain high-precision numerical results at arbitrary points in phase-space by joining series expansions along multiple connected line segments. The main integration strategy behind the DiffExp package builds on ideas originating from Ref. \cite{Francesco:2019yqt}, which in turn builds on a large set of previous literature on series expansions methods for Feynman integrals.

For many physical amplitudes in QCD or the Standard Model, the main computational bottlenecks are the integration-by-parts (IBP) reduction of large sets of scalar Feynman integrals to a basis of linearly independent master integrals, and the efficient evaluation of these master integrals in the physical region of the given process. Many computations in phenomenology are easy to perform at leading order in the coupling constants, much more difficult at next-to-leading-order, and can be near impossible with current techniques at next-to-next-to leading order, unless suitable approximations are employed in the theory. The procedure of IBP-reduction has been given a lot of attention in recent years, and numerous specialized packages based on the Laporta \cite{Laporta:2001dd} algorithm have been developed, such as \texttt{LiteRed}, \texttt{Fire}, and \texttt{Kira} \cite{Lee:2013mka, Smirnov:2019qkx, Klappert:2020nbg}. Given sufficiently complicated processes or high loop orders, these packages will run into computational limits, but there are many cases where the ability to efficiently evaluate the master integrals lags behind the ability to perform the reductions.

Feynman integrals may be computed using numerical methods, such as Monte-Carlo integrators combined with sector decomposition techniques (see e.g. \cite{Smirnov:2015mct, Borowka:2017idc}), or using analytic methods. When available, analytic methods are generally much faster than Monte-Carlo based integration methods, but it is not always known how to obtain a given Feynman integral in an analytic way. Let us briefly review some of the available analytic methods. Firstly, it is well-known that many simple Feynman integrals admit representations in terms of combinations of hypergeometric functions, in closed form in the dimensional regulator $\epsilon$. In recent works, it has also been shown that A-hypergeometric systems (also called GKZ hypergeometric systems) are sufficient to describe more complicated families of Feynman integrals (see e.g. \cite{Vanhove:2018mto, delaCruz:2019skx, Klausen:2019hrg, Feng:2019bdx, Klemm:2019dbm, Bonisch:2020qmm}).  The Feynman integrals are then expressed as sums over multidimensional canonical series. Because these series have a limited range of convergence, it can in practice be difficult to evaluate Feynman integrals at arbitrary points in phase-space using this approach. However, in principle this can be done, and for example in Ref. \cite{Bonisch:2020qmm} a Pari program is presented for the evaluation of the equal-mass four-loop banana graph in integer dimension $d=2$.

For phenomenology, one is usually interested in computing the Laurent series of the integrals up to some order in the dimensional regulator $\epsilon$. When calculating Feynman integrals this way, it is typically possible to express them in terms of iterated integrals of integration kernels defined on some geometrical space. Two powerful methods for obtaining such results are the method of direct integration of the Feynman parametrization (see e.g. \cite{Brown:2009ta, Panzer:2014caa}), and differential equation methods. Roughly, the complexity of the kernels appearing in the iterated integrals seems to be dictated by the maximal cut of the Feynman integral, and the maximal cuts of its subtopologies (see also Ref. \cite{Primo:2016ebd} for connections between the maximal cut and differential equations.) When all maximal cuts are rational and/or algebraic it is often possible to express a Feynman integral in terms of multiple polylogarithms \cite{Goncharov:1998kja}, while the presence of maximal cuts that evaluate to elliptic integrals indicates that more complicated functions are needed, such as elliptic multiple polylogarithms (see e.g. \cite{brown2011multiple, Adams:2014vja, Broedel:2014vla, Broedel:2017kkb, Broedel:2017siw, Adams:2018yfj, Bogner:2019lfa}.)

It is expected that kernels defined on increasingly complicated geometries have to be considered as the number of loops and scales are increased, such as those involving hyperelliptic curves \cite{Zhang:2016hdb}, or Calabi-Yau geometries. For example, the so-called "banana" graphs are associated with Calabi-Yau $(l-1)$-folds, where $l$ denotes the number of loops (see e.g. \cite{Vanhove:2018mto, Klemm:2019dbm}). Unfortunately, the iterated integrals for geometries beyond elliptic curves have not yet been studied in detail. Furthermore, even in cases where the geometry of the maximal cuts seems simple, it might not be known or even possible to evaluate the integrals in terms of multiple polylogarithms or elliptic polylogarithms. For example, some families of integrals admit a canonical $d\log$-basis which depends on numerous non-simultaneously rationalizable square roots. In such a case it can be very difficult to obtain polylogarithmic expressions (which admit fast evaluation) for the integrals - especially at higher weights - and it might not be possible to obtain polylogarithmic expressions at all (see e.g. Refs. \cite{Besier:2018jen, Besier:2019kco, Heller:2019gkq, Besier:2020hjf, Brown:2020rda} for recent works.) Similarly, when multiple square roots are coupled to elliptic sectors, it is not clear how to express the elliptic integrals in terms of elliptic multiple polylogarithms

This situation arises for the master integrals relevant for Higgs plus jet production at next-to-leading order in QCD with full heavy quark mass dependence \cite{Bonciani:2016qxi, Bonciani:2019jyb, Frellesvig:2019byn}. The presence of an internal mass introduces numerous square roots in the definition of the canonical basis of the polylogarithmic sectors, and renders the top sectors elliptic. The reductions of the planar families of integrals were performed in Ref. \cite{Bonciani:2016qxi}, and an analytic representation for the integrals was obtained in that paper in terms of integrals over polylogarithms of weight 2, and elliptic integration kernels. While the resulting expressions solve the integrals in principle, it can be slow to obtain high precision numerical results from them, and difficult to perform the analytic continuation of the results to the physical region. 
In Ref. \cite{Francesco:2019yqt}, a powerful strategy was introduced for evaluating the planar master integrals of Ref. \cite{Bonciani:2016qxi}, and for analytically continuing them to the physical region for Higgs plus jet production. Furthermore, in Refs. \cite{Bonciani:2019jyb} and \cite{Frellesvig:2019byn}, the computation of the non-planar master integrals was performed using the same methods. The integration method relies on iteratively solving the integrals from their differential equations in terms of truncated one-dimensional series expansions along line segments in phase-space. The series expansions may be truncated at relatively high orders (usually 50 to 100) in order to obtain results in the tens of digits, which is generally more than sufficient for phenomenological computations. Furthermore, when the series expansions are centered at branch points, they contain logarithms and square roots. The analytic continuation of these elementary functions completely describes the analytic continuation of the Feynman integrals, which trivializes the procedure of analytic continuation. 

The Mathematica package that is the subject of this paper aims to provide a general purpose public implementation of the series expansion methods considered in Refs. \cite{Francesco:2019yqt, Bonciani:2019jyb, Frellesvig:2019byn, Abreu:2020jxa}. The rest of this paper is organized as follows. In Section \ref{sec:feynmanintegralreview}, we review scalar Feynman integrals, and introduce some of the notation used throughout this paper. We also review how to obtain results at asymptotic limits using the method of expansions by regions in the Feynman parametrization. In Section \ref{sec:differentialequations}, we review the method of differential equations. In Section \ref{sec:seriesexpansionmethods}, we discuss all aspects of solving Feynman integrals as one-dimensional series expansions from the differential equations. In particular, we discuss how to obtain an integration sequence, how to solve the homogeneous and inhomogeneous differential equations of coupled Feynman integrals, we discuss how to perform the analytic continuation past threshold singularities and branch points, we discuss how to improve the precision of the series expansions, and lastly we discuss strategies for obtaining the line segments along which to integrate. In Section \ref{sec:DiffExp}, we discuss the main functions and options of the DiffExp package. Lastly, in Section \ref{sec:examples} we discuss the computation of the equal-mass and unequal-mass three-loop banana graph families with DiffExp. We also apply DiffExp to a few examples from the literature. The conclusions to the paper are given in Section \ref{sec:conclusions}.

\section{Review of some aspects of Feynman integrals}
\label{sec:feynmanintegralreview}
In the following section, we review some basic properties of scalar Feynman integrals.
\subsection{Basic definitions}
\label{sec:fimain}
Suppose we are given a Feynman diagram, for which we denote the number of loops by $l$, and the number of propagators by $n$. We may then define a family of scalar Feynman integrals associated with the Feynman diagram as a collection of integrals of the form
\begin{align}
    \label{eq:FIMomRep}
    I_{a_1,\ldots,a_{n+m}} = \int\left(\prod_{i=1}^{l} d^{d} k_{i}\right) \frac{\prod_{i=n+1}^{n+m} N_i^{-a_i}}{\prod_{i=1}^n D_{i}^{a_i}}\,,\quad D_i = -q_{i}^2 + m_{i}^2 -i\delta\,,
\end{align}
where we take the indices $a_i$ to be integers, of which $a_1,\ldots,a_n$ are non-negative, and of which $a_{n+1},\ldots,a_{n+m}$ are non-positive. Each propagator $D_i$ inherits its internal momentum $q_i$ from the Feynman diagram. The factors $i\delta$, with $\delta>0$ being an infinitesimally small positive number, are introduced as part of the Feynman prescription and invoke a (physical) choice of branch of the Feynman integrals. We elaborate more on this point in Section \ref{sec:analyticcontinuationremarks}. The numerator terms $N_i$ are linear combinations of dot products of internal and external momenta, and can be freely chosen subject to the constraint that the propagators and numerators form a basis of the vector space of dot products of the form $k_i\cdot k_j$ and $k_i \cdot p_j$, where $k_i$ denotes a loop momentum, and where $p_j$ denotes an external momentum.

It is well-known that integrals within a family may be related to each other through IBP-identities. In particular, it is possible to express any member of a family of Feynman integrals as a linear combination of a finite basis of linearly independent Feynman integrals in the given family. The choice of independent basis is called a choice of master integrals. It is often possible to choose a basis of master integrals without numerators. Feynman integrals are often divergent, and have to be computed through a suitable regularization prescription. A powerful regularization prescription is dimensional regularization. In dimensional regularization, the dimension $d$ is upgraded to a complex parameter, usually written as $d_{0} - 2\epsilon$, where $d_{0}$ is an integer, and where $\epsilon$ is called the dimensional regulator. This does not immediately make sense from the viewpoint of Eq. (\ref{eq:FIMomRep}), but it can be made rigorous by first converting Eq. (\ref{eq:FIMomRep}) to a parametric representation such as the Feynman parametrization (see Section \ref{sec:feynpar}), in which the dimension $d$ becomes a variable in the integrand that is roughly on the same footing as the powers of the propagators. The (infrared and ultraviolet) divergences of the Feynman integral are then expressed as poles in the dimensional regulator.

Feynman integrals satisfy the scaling relation:
\begin{equation}
\label{eq:eqnscaling}
I_{a_1, \ldots, a_{n+m}}(S/\lambda)=\lambda^{-\frac{\gamma}{2}} I_{a_1, \ldots, a_{n+m}}(S)\,,\quad \gamma=ld-2\sum_j a_j\,,
\end{equation}
where we explicitly wrote the dependence on the set $S = \{p_{j}^2\} \cup \{s_{ij}\} \cup \{m_j^2\}$, containing the squares of external momenta, the Mandelstam variables, and the internal masses, and where by $S/\lambda$ we denote the set of elements $\{s/\lambda\,|\, s \in S\}$, where $\lambda$ is a parameter of mass dimension two. Furthermore, note that $\gamma$ is the mass dimension of the integral. By choosing $\lambda\in S$, we may trivialize the dependence on one of the kinematic invariants or internal masses.

\subsection{Feynman parametrization}
\label{sec:feynpar}
Often it is useful to rewrite Feynman integrals in a parametric representation such as the Feynman parametrization. Although this paper deals with differential equation methods, the Feynman parametrization is still of use for computing boundary conditions in asymptotic limits, which we will discuss in Section \ref{sec:expansionbyregions}. A Feynman integral for which all the numerators have exponent zero, admits the following Feynman parametrization:
\begin{align}
    \label{eq:feynpar}
    I_{a_1, \ldots, a_{n}}=\left(i \pi^{\frac{d}{2}}\right)^{l}\Gamma\left(a-\frac{l d}{2}\right)\int_{\Delta^{n-1}} [d^{n-1} \vec{\alpha}]   \left(\prod_{i=1}^{n} \frac{\alpha_{i}^{a_{i}-1}}{\Gamma(a_i)}\right) \mathcal{U}^{a-\frac{d}{2}(l+1)} \mathcal{F}^{-a+\frac{l d}{2}}
\end{align}
where $a = a_1 + \ldots + a_n$, where $\Delta^{n-1}=\left\{\left[\alpha_{1}: \alpha_{2}: \ldots: \alpha_{n}\right] \in \mathbb{R}\mathbb{P}^{n-1} \mid \alpha_{i} \geq 0,1 \leq i \leq n\right\}$, and where $[d^{n-1}\vec{\alpha}]$ denotes the canonical volume form on $\mathbb{R} \mathbb{P}^{n-1}$, given by:
\begin{align}
    \left[d^{n-1} \vec{\alpha}\right]\equiv \sum_{j=1}^{n}(-1)^{j-1} \alpha_{j} d \alpha_{1} \wedge \cdots \wedge \widehat{d \alpha}_{j} \wedge \cdots \wedge d \alpha_{n}\,.
\end{align}
The so-called Symanzik polynomials $\mathcal{U}$ and $\mathcal{F}$ can be written in terms of the Feynman diagram $G$ as:
\begin{align}
\label{eq:symanzikpols}
\mathcal{U}=\sum_{T \in T(G)} \prod_{e_{i} \notin T} \alpha_{i}, \,\, \tilde{\mathcal{F}}=\sum_{\left(T_{1}, T_{2}\right) \in F(G)}\left(\prod_{e_{i} \notin\left(T_{1} \cup T_{2}\right)} \alpha_{i}\right) s_{\left(T_{1}, T_{2}\right)}, \,\, \mathcal{F}=-\tilde{\mathcal{F}}+\mathcal{U}\left(\sum \alpha_{i} m_{i}^{2}\right)\,,
\end{align}
where $T(G)$ denotes the set of spanning trees of $G$, and where $F(G)$ denotes the set of all two-forest of $G$. Note that a two-forest is a set of two disjoint trees whose union touches all the vertices of the graph. We denoted the square of the momentum flowing between the components $T_1$ and $T_2$ by $S_{(T_1,T_2)}$.

The integration variables $\alpha_j$ are referred to as Feynman parameters. The Cheng-Wu theorem \cite{chengwu} tells us that by a change of variables we may pull-back the projective integration to a simplex:
\begin{align}
    \label{eq:projtosimplex}
    \int_{\Delta^{n-1}}\left[d^{n-1} \vec{\alpha}\right] \rightarrow \int_{\mathbb{R}_{\geq 0}^{n}} d^{n} \vec{\alpha}\, \delta\left(1-\sum_{j = 1}^n \alpha_{j}\right)\,,
\end{align}
where $J\subseteq [1,n]$ may be chosen to be any nonempty subset of the Feynman parameters, and where we choose the orientation of the integration over $\Delta^{n-1}$ that gives a positive sign on the right-hand side of the equation.

In the case where the Feynman integral has numerators (i.e. some of the $a_i$ are negative integers for $i > n$), we need to do a bit more work to give the Feynman parametrization. First, we need a definition of the Symanzik polynomials that derives directly from the propagators $D_i$. Consider the $(l\times l)$-matrix $A$, $l$-vector $B$, and constant $C$, defined by:
\begin{align}
    \sum_{i=1}^{n} \alpha_{i} D_i +    \sum_{i=n+1}^{n+m} \alpha_{i} N_i  =-\sum_{i,j=1}^{l} k_{i} A_{ij} k_{j}+\sum_{i=1}^{l} 2 k_{i} \cdot B_{i}+C\,.
\end{align}
We then let:
\begin{align}
    \mathcal{U}^+=\det(A)\,, \quad \mathcal{F}^+=\det(A)\left(C+B A^{-1} B\right)\,.
\end{align}
The Feynman parametrization is then given by:
\begin{align}
    I_{a_1, \ldots, a_{n+m}}&=\left(i \pi^{\frac{d}{2}}\right)^{l}\Gamma\left(a-\frac{l d}{2}\right)\int_{\Delta^{n-1}} [d^{n-1} \vec{\alpha}]   \left(\prod_{i=1}^{n} \frac{\alpha_{i}^{a_{i}-1}}{\Gamma(a_i)}\right) \left[\left(\prod_{j={n+1}}^{n+m}(-1)^{a_j}\frac{\partial^{-a_i}}{\partial \alpha_j^{-a_i}}\right)\times\right.\nonumber\\
    &\quad\quad\left(\mathcal{U}^+\right)^{a-\frac{d}{2}(l+1)} \left(\mathcal{F}^+\right)^{-a+\frac{l d}{2}}\bigg]\bigg|_{\alpha_{n+1},\ldots,\alpha_{n+m} = 0}\,,
\end{align}
where $a = a_1 + \ldots + a_{n+m}$\,. Note that $\mathcal{U}^+|_{\alpha_{n+1},\ldots,\alpha_{n+m} = 0} = \mathcal{U}$, and that $\mathcal{F}^+|_{\alpha_{n+1},\ldots,\alpha_{n+m} = 0} = \mathcal{F}$. See also Ref. \cite{Bogner:2010kv} for a more detailed review of Feynman graph polynomials.

\subsection{Remarks on analytic continuation}
\label{sec:analyticcontinuationremarks}
In this section, we will remark upon a few aspects of the analytic continuation of Feynman integrals. Let us consider how the Feynman prescription in the momentum space representation translates to the Feynman parametrization. First, we may absorb the $i\delta$'s in the definition of the internal masses. Then, looking at Eq. (\ref{eq:symanzikpols}), we see that:
\begin{align}
    \mathcal{F} \rightarrow \mathcal{F} - i\delta\, \mathcal{U}\,.
\end{align}
and since $\mathcal{U}$ is positive-definite we can put:
\begin{align}
    \label{eq:Fidelta}
    \mathcal{F} \rightarrow \mathcal{F} - i\delta\,.
\end{align}
Therefore, the second Symanzik polynomial $\mathcal{F}$ carries an infinitesimally small negative imaginary part. The integration of the Feynman parametrization is the simplest in a region where $\mathcal{F} > 0$ on the interior of the whole integration domain, as the $i\delta$ prescription can then be dropped. From Eq. (\ref{eq:symanzikpols}) we see that letting $s_{(T_1,T_2)}<0$ for all two-forests is sufficient for this condition to hold. This kinematic region is known as the Euclidean region. Note that such a region is not always guaranteed to exist. For an example see e.g. Section 4.1 of Ref. \cite{Henn:2013nsa}.

In the Euclidean region, the only possible singularities of the Feynman integral lie at the boundary of the integration domain, where we may have $\mathcal{U} = 0$, or $\mathcal{F} = 0$.  If we choose the integration domain to be a simplex containing all Feynman parameters, i.e. the set $\{(\alpha_1,\ldots,\alpha_n)\,|\,\alpha_i \geq 0,\,\alpha_1 + \ldots \alpha_n = 1\}$, then all possible boundary singularities lie at positions where subsets of the Feynman parameters vanish. If we apply the Cheng-Wu theorem and choose a different integration domain, for example the set $\{(\alpha_1,\ldots,\alpha_n)\,|\, \alpha_i \geq 0,\, \alpha_n = 1\}$, then there may also be singularities when subsets of integration variables go out to infinity. Using the method of analytic regularization\footnote{Not to be confused with the identically named concept of analytic regularization in the method of expansion by regions, where the propagator exponents are used as additional regulators.} from Ref. \cite{Panzer:2014gra}, it is possible to rewrite a Feynman integral in the Feynman parametrization in terms of a sum of integrals with prefactors that depend on $\epsilon$, for which there are no more boundary singularities in the integration domain. The terms in the sum are Feynman integrals associated with the same graph, but with different propagator powers and shifted dimensions. This method is implemented in the package HyperInt \cite{Panzer:2014caa}. Another approach to resolve boundary singularities is the method of sector decomposition \cite{Binoth:2000ps, Binoth:2003ak, Bogner:2007cr}.

Outside of the Euclidean region, Feynman integrals have threshold singularities. The locations of these singularities can be found from the Landau equations \cite{Landau:1959fi}, which we will not discuss here further. Instead of integrating Feynman integrals directly in a given physical region, it is usually simplest to first perform the integration in the Euclidean region, and to analytically continue to the physical region from there. It is important that threshold singularities are crossed in a manner that is consistent with the Feynman prescription. Looking at Eqns. (\ref{eq:symanzikpols}) and (\ref{eq:Fidelta}), we see that every squared mass should be interpreted to carry a negative imaginary part, while the Mandelstam variables $s_{(T_1,T_2)}$ should carry a positive imaginary part, since their prefactors in the second Symanzik polynomials are sums of monomials with positive coefficients.

\subsection{Expansions around asymptotic limits}
\label{sec:expansionbyregions}
In Section \ref{sec:seriesexpansionmethods}, we will describe the method of differential equations for Feynman integrals. To fix the solution of a family of Feynman integrals from differential equations, it is important to find boundary conditions at a suitable point or limit. In this section we briefly review the method of expansion by regions \cite{Beneke:1997zp, Smirnov:1999bza, Jantzen:2011nz}, which may be used to find boundary conditions in asymptotic limits. (See also Refs. \cite{Semenova:2018cwy, Ananthanarayan:2018tog, Ananthanarayan:2020ptw} for some recent developments.) 

Generally, we would like to compute boundary conditions in special points, where the Feynman integrals are expected to simplify. If one looks naively at the second Symanzik polynomial, it seems that the simplest choice of boundary point should be one where most of the kinematic invariants and internal masses vanish. In such a point, the second Symanzik polynomial will simplify, and the Feynman parametrization may then often be integrated in closed form in $\epsilon$ in terms of simple functions, such as ratios of gamma functions. However, typically a Feynman integral develops divergences as we approach such a point, and we would not obtain the correct asymptotic limit by simply plugging it into the integrand. To illustrate this with a simple example, let us consider the massive bubble, dimensionally regulated around $d=2-2\epsilon$. We have:
\begin{align}
\label{eq:masbub}
\frac{e^{\gamma_E \epsilon}}{i \pi^{1-\epsilon}}\int d^{d} k_{1} \frac{1}{(-k_1^2+m^2)(-(k_1+p)^2+m^2)} = \frac{2 \log \left(\frac{-\sqrt{-p^{2}}-\sqrt{4 m^{2}-p^{2}}}{\sqrt{-p^{2}}-\sqrt{4 m^{2}-p^{2}}}\right)}{\sqrt{-p^{2}} \sqrt{4 m^{2}-p^{2}}} + \mathcal{O}(\epsilon)\,,
\end{align}
in the Euclidean region. Note that the factor $e^{\gamma_E \epsilon}/i \pi^{1-\epsilon}$ was added by convention, where $\gamma_E$ is the Euler-Mascheroni constant. Next, let us consider the zero-mass limit. In particular, we let $m^2 = x$, and we take the limit $x \downarrow 0$. This yields the following expression at finite order in $\epsilon$:
\begin{align}
    \label{eq:bubasy}
    -\frac{2\left(\log \left(-p^2\right)-\log (x )\right)}{p^{2}}+\mathcal{O}(x)\,.
\end{align}
If we start directly from the massless bubble, we find instead:
\begin{align}
    \label{eq:masslbub}
    \frac{e^{\gamma_E \epsilon}}{i \pi^{1-\epsilon}}\int d^{d} k_{1} \frac{1}{(-k_1^2)(-(k_1+p)^2)} = \frac{2}{p^{2} \epsilon}-\frac{2 \log \left(-p^{2}\right)}{p^{2}}+\mathcal{O}(\epsilon)\,.
\end{align}
Thus, the kinematic singularity in the asymptotic limit in Eq. (\ref{eq:bubasy}) shows up as a dimensionally regulated singularity in Eq. (\ref{eq:masslbub}), and we can not use Eq. (\ref{eq:masslbub}) to provide the boundary conditions for the massive bubble in the massless limit. The question is then how to obtain the asymptotic limit without first computing the integral for a generic configuration of $p^2$ and $m^2$, which defeats the purpose of choosing a simple boundary point. One solution is to use the method of expansion by regions \cite{Beneke:1997zp}. The method has a powerful formulation in the Feynman parametrization, which was developed in Refs. \cite{Pak:2010pt, Jantzen:2012mw}. Furthermore, Ref. \cite{Jantzen:2012mw} comes with a powerful Mathematica package \texttt{asy}, that implements the method.\footnote{Note that the latest version of \texttt{asy.m} is shipped together with the program FIESTA \cite{Smirnov:2015mct}.}

We briefly outline the method next, from a pragmatic viewpoint. Suppose we consider a Feynman integral with $n$ propagators, which depends on a set of kinematic invariants and internal masses $S = \{s_1,\ldots,s_{|S|}\}$ where $|S|\geq 1$, and which is written in the Feynman parametrization. Next, suppose that we are interested in obtaining the asymptotic behaviour in a one-scale limit in which every kinematic invariant and mass has a certain scaling $s_i \rightarrow s_i' = x^{\gamma_i}s_i$ for $i = 1,\ldots,|S|$, where the exponents $\gamma_i$ are rational numbers, and where $x$ is a line parameter that goes to zero. The method of expansion by regions states that there is a set of regions $\{R_i\}$, denoted by $R_i = (r_{i1},\ldots,r_{in})$ for each $i$, which describe rescalings of the Feynman parameters, and which prescribe how to compute the asymptotic expansion in the limit. The set of regions can be determined from the Symanzik polynomials of the Feynman integral and also depends on the asymptotic limit that is being considered.

We will not discuss the derivation of the set of regions here. We note that they can be obtained using for example the program \texttt{asy.m}, which relies on a geometric algorithm based on finding the convex hull of a set of points determined from the Symanzik polynomials \cite{Jantzen:2012mw}. For each region, we rescale the Feynman parameters and their differentials according to $\alpha_j \rightarrow \alpha_j' = x^{r_{ij}}\alpha_j$. In addition, we also rescale the kinematic parameters and masses according to $s_i \rightarrow s_i' = x^{\gamma_i}s_i$. Next, we expand the contribution of each region in the line parameter $x$, we integrate the result, and we sum the contributions together. The claim of the method of expansion by regions is that the resulting sum provides the asymptotic limit of the Feynman integral. Note that it is currently not fully mathematically proven that the method is correct \cite{Semenova:2018cwy}, however in practice the method is known to work from the consideration of many examples.

Let us reconsider the example of the massive bubble. Its Feynman parametrization in $d=2-2\epsilon$ is given by:
\begin{align}
    e^{\gamma_E \epsilon} \Gamma(\epsilon+1) \int_{\Delta^{1}}\left[d^{1} \vec{\alpha}\right] \left(\alpha_{1}+\alpha_{2}\right)^{2 \epsilon} \left(\alpha_{1}^{2} m^{2}+\alpha_{2}^{2} m^{2}+2 \alpha_{1} \alpha_{2} m^{2}-\alpha_{1} \alpha_{2} p^{2}\right)^{-1-\epsilon}\,.
\end{align}
Using the Mathematica package \texttt{asy} we obtain the regions
\begin{align}
    R_1 = \{0,0\},\quad R_2 = \{0,-1\},\quad  R_3 = \{0,1\}\,,
\end{align}
in the asymptotic limit $m^2 = x \downarrow 0$. Rescaling the Feynman parameters in each region, and summing over the result yields the expression:
\begin{align}
    &e^{\gamma_E \epsilon} \Gamma(\epsilon+1)\int_{\Delta^{1}}\left[d^{1} \vec{\alpha}\right]\bigg(
   \left(\alpha_{1}+\alpha_{2}\right)^{2 \epsilon}\left(  x  \alpha_{1}^{2}-p^{2} \alpha_{1} \alpha_{2}+2  x  \alpha_{1} \alpha_{2}+ x \alpha_{2}^{2}\right)^{-1-\epsilon} \nonumber\\
    &\quad+ x^{-\epsilon}\left(x  \alpha_{1}+\alpha_{2}\right)^{2 \epsilon}\left( x^{2}  \alpha_{1}^{2}-p^{2} \alpha_{1} \alpha_{2}+2  x  \alpha_{1} \alpha_{2}+ \alpha_{2}^{2}\right)^{-1-\epsilon} \nonumber\\
    &\quad+x^{-\epsilon}\left(\alpha_{1}+x  \alpha_{2}\right)^{2 \epsilon}\left( \alpha_{1}^{2}-p^{2} \alpha_{1} \alpha_{2}+2  x \alpha_{1} \alpha_{2}+ x^{2}  \alpha_{2}^{2}\right)^{-1-\epsilon}\bigg)\,.
\end{align}
At leading order in $x$ we obtain:
\begin{align}
    &e^{\gamma_E \epsilon} \Gamma(\epsilon+1)\int_{\Delta^{1}}\left[d^{1} \vec{\alpha}\right]\bigg(\alpha_{1}^{-\epsilon-1} \alpha_{2}^{-\epsilon-1}\left(\alpha_{1}+\alpha_{2}\right)^{2 \epsilon}\left(-p^{2}\right)^{-1-\epsilon} \nonumber\\
    &\quad+x^{-\epsilon}\alpha_{2}^{-1+\epsilon}\left(-p^{2} \alpha_{1}+ \alpha_{2}\right)^{-1-\epsilon}+ x^{-\epsilon}\alpha_{1}^{\epsilon-1}\left( \alpha_{1} -p^{2}\alpha_{2} \right)^{-\epsilon-1}\bigg)\,.
\end{align}
After integrating the result, we find:
\begin{align}
   -e^{\gamma_E \epsilon} \frac{\Gamma(\epsilon)}{p^2}\left(\epsilon\frac{\left(-p^{2}\right)^{-\epsilon} \Gamma(-\epsilon)^{2} }{\Gamma(-2 \epsilon)}+2  x^{-\epsilon} \right) = -\frac{2\left(\log \left(-p^{2}\right)-\log ( x)\right)}{p^{2}}+\mathcal{O}(\epsilon)\,,
\end{align}
which agrees with Eq. (\ref{eq:bubasy}). In Section \ref{sec:examples}, we will use the method of expansion by regions to obtain boundary terms for the three-loop banana graphs.

\section{The differential equations method}
\label{sec:differentialequations}
In the following section we review the method of differential equations for Feynman integrals. 

\subsection{Basic definitions}
An important property of Feynman integrals is that they can be realized as solutions to linear systems of ordinary differential equations with respect to the kinematic invariants and internal masses \cite{Kotikov:1990kg, Kotikov:1991pm, Kotikov:1991hm}. The traditional way to see this, is to take a basis of master integrals of a given family and to note that their derivatives can be expressed as combinations of Feynman integrals in the same family with different propagator exponents. These integrals may be IBP-reduced back to the original set of master integrals, which allows one to write the derivatives for the master integrals in terms of a closed-form linear system of differential equations. In the following we review a few basic properties of these differential equations.

Let us consider a family of Feynman integrals with $m$ master integrals, packaged into a vector $\vec{f} = (f_1,\ldots,f_m)$. Suppose that the Feynman integrals depend on a set of kinematic invariants and internal masses that we denote by $S$, which consist of squares of sums of external momenta, and of squares of internal masses. We may write the associated system of differential equations in the following form:
\begin{align}
    \label{eq:basicdeqns}
    d \vec{f} = \left(\sum_{s\in S}\mathbf{A}_s\,ds\right) \vec{f}\,,
\end{align}
where we will refer to the matrices $\mathbf{A}_s$ as partial derivative matrices. From the vanishing of the total differential $d^2 = 0$, we have the integrability condition:
\begin{align}
    \label{eq:deqnintegrability}
    \partial_{s_1} \mathbf{A}_{s_2} - \partial_{s_2} \mathbf{A}_{s_1} + [\mathbf{A}_{s_1},\mathbf{A}_{s_2}] = 0\quad\text{for all }s_1,s_2\in S\,.
\end{align}
If we have $d\tilde{\mathbf{A}} = \sum_{s \in S} \mathbf{A}_{s} d s$, then we may also write the above equation as:
\begin{equation}
    d\tilde{\mathbf{A}}=\tilde{\mathbf{A}} \wedge \tilde{\mathbf{A}}\,.
\end{equation}
Another property of the differential equations is the scaling relation. Starting from Eq. (\ref{eq:eqnscaling}), taking a derivative with respect to $\lambda$, and putting $\lambda = 1$ yields:
\begin{align}
    \sum_{s \in S} s\partial_s I_{ a_1, \ldots, a_{n+m}}=\frac{\gamma}{2} I
    _{ a_1, \ldots, a_{n+m}}\,,
\end{align}
where $S$ is the set of kinematic invariants and internal masses. This in turn leads to:
\begin{align}
    \sum_{s \in S} s \mathbf{A}_s = \Gamma\,,
\end{align}
where $\Gamma$ is the diagonal matrix with entries $\gamma_j/2$, where $\gamma_j$ denotes the mass dimension of the $j$-th basis integral. It is often a good idea to verify that the integrability condition and the scaling relation are satisfied as a cross-check that the differential equations were derived correctly.

\subsection{Canonical basis}
\label{sec:canonicalbases}
The differential equations may be considerably simplified when a so-called canonical choice of basis is made, a concept that was introduced in Ref. \cite{Henn:2013pwa}. Let us first consider a generic change of basis, $\vec{B} = \mathbf{T}^{-1} \vec{f}$, where $\mathbf{T}$ is some matrix that may depend on the kinematic invariants, on the internal masses, and on $\epsilon$. The partial derivative with respect to a variable $s$ then takes the form:
\begin{align}
    \label{eq:canformpartials1}
    \frac{\partial}{\partial s}\vec{B} = \left[\left(\partial_{s}\mathbf{T}^{-1}\right)\mathbf{T}+\mathbf{T}^{-1}\mathbf{A}_{s}\mathbf{T}\right] \vec{B}\,.
\end{align}
It was observed in Ref. \cite{Henn:2013pwa} that if $\mathbf{T}$ is chosen such that
\begin{align}
    \label{eq:changeofbasiscan}
    \left(\partial_{s}\mathbf{T}^{-1}\right)\mathbf{T}+\mathbf{T}^{-1}\mathbf{A}_{s}\mathbf{T} = \epsilon \tilde{\mathbf{A}}_s\,, 
\end{align}
for all kinematic invariants and internal masses $s \in S$, and where $\tilde{\mathbf{A}}_s$ is independent of $\epsilon$, the differential equations are simplified considerably. It was furthermore conjectured in Ref. \cite{Henn:2013pwa} that there is always such a choice of matrix $\mathbf{T}$.

For integrals that are expressible in terms of multiple polylogarithms, the canonical basis may be written in the form:
\begin{align}
    \label{eq:canformpoly}
    d \vec{B} = \epsilon d\tilde{\mathbf{\mathbf{A}}} \vec{B}\,,\quad \tilde{\mathbf{A}} = \sum_{l \in \mathcal{A}} \tilde{\mathbf{A}}_l \log(l) \,,
\end{align}
where $\tilde{\mathbf{A}}_l$ are matrices of rational numbers, and where $\mathcal{A}$ is a set of functions of the kinematic invariants and internal masses, called the alphabet, whose elements are called letters. Note that in the mathematics literature, the alphabet usually denotes instead the set of differential one-forms $d\log(l)$.

The general solution to Eq. (\ref{eq:canformpoly}) may be written in terms of a path-ordered exponential:
\begin{align}
    \vec{B}=\mathbb{P} \exp \left( \int_{\gamma}  \epsilon\, d\tilde{\mathbf{A}} \right) \vec{B}(\gamma(0))\,,
\end{align}
where $\gamma: [0,1] \rightarrow \mathbb{C}^{|S|}$ is a path in the phase-space of the kinematic invariants and internal masses $S$, and where $|S|$ denotes the number of these. Let us denote the expansion in $\epsilon$ of the basis integrals by:
\begin{align}
    \vec{B} = \sum_{k=0}^\infty \vec{B}^{(k)} \epsilon^k\,,
\end{align}
where we assume the expansion starts at finite order in $\epsilon$. Note that this can always be achieved by multiplying the basis by an overall power of $\epsilon$. Expanded in terms of iterated integrals, the path-ordered exponential works out to:
\begin{align}
\vec{B}=\vec{B}^{(0)}(\gamma(0))+\sum_{k \geq 1} \epsilon^{k} \sum_{j=1}^{k} \int_{0}^{1} \gamma^{*}(d \tilde{\mathbf{A}})\left(t_{1}\right) \int_{0}^{t_{1}} \gamma^{*}(d \tilde{\mathbf{A}})\left(t_{2}\right) \ldots \int_{0}^{t_{j-1}} \gamma^{*}(d \tilde{\mathbf{A}})\left(t_{j}\right) \vec{B}^{(k-j)}(\gamma(0))
\end{align}
To obtain a matrix $\mathbf{T}$ that solves Eq. (\ref{eq:changeofbasiscan}), it is useful to first find a precanonical basis, in which the differential equations are given by:
\begin{align}
    \frac{\partial}{\partial s} \vec{f} = \left(\mathbf{A}_s^{(0)} + \epsilon \mathbf{A}_s^{(1)}\right) \vec{f}\,,
\end{align}
for all kinematic invariants and internal masses $s$, and where the matrices $\mathbf{A}_s^{(0)}$ and $\mathbf{A}_s^{(1)}$ do not depend on $\epsilon$. Such a precanonical basis may often be found by performing a change of basis where the prefactors depend on $\epsilon$ but not on the kinematic invariants and internal masses. If we start from a precanonical basis, Eq. (\ref{eq:changeofbasiscan}) is solved by a matrix $\mathbf{T}$ that is independent of $\epsilon$, and which satisfies:
\begin{align}
    \partial_{s}\mathbf{T} = \mathbf{A}_{s}^{(0)}\mathbf{T}\,.
\end{align}
Hence, $\mathbf{T}$ is an invertible matrix that satisfies the precanonical differential equations at leading order.

\section{Series expansion methods}
\label{sec:seriesexpansionmethods}
In this section, we will outline how to find series solutions for Feynman integrals starting from their systems of differential equations. The core ideas are based on the integration strategy of Ref. \cite{Francesco:2019yqt}, which was further studied and applied in Refs. \cite{Frellesvig:2019byn, Bonciani:2019jyb} in the context of Higgs plus jet integrals. The strategy has also been applied recently to the computation of two-loop non-planar five-point functions in Ref \cite{Abreu:2020jxa}. Series expansions methods have also been explored in many other literature, such as in Refs. \cite{Pozzorini:2005ff,Aglietti:2007as,Mueller:2015lrx,Melnikov:2016qoc,Lee:2017qql,Melnikov:2017pgf,Lee:2018ojn,Bonciani:2018uvv,Mistlberger:2018etf,Bonciani:2018omm, Bruser:2018jnc,Davies:2018ood,Davies:2018qvx,Heller:2019gkq}, usually for the computation of single-scale integrals, or for the computation of multi-scale integrals in special kinematic limits.

In the series expansion method, one considers multiple one-dimensional series expansions along a set of connected line segments, which start from a given boundary point and end up at the desired point in phase-space. It is necessary to consider multiple expansions, because the series expansions on each line segment only converge within a certain radius. Branch points and singularities may be crossed by centering a line segment at the branch point or singularity. The series solutions to the differential equations may then contain square roots and logarithms. The analytic continuation of these functions can be performed by assigning an imaginary part of the form $\pm i\delta$ to the line parameter, in accordance with the Feynman prescription.

We provide a few improvements here compared to Refs. \cite{Francesco:2019yqt, Frellesvig:2019byn, Bonciani:2019jyb}. In Section \ref{sec:intorder} we discuss how to derive an integration sequence directly from the differential equations. In Section \ref{sec:homogeneoussolutions} we discuss a simple way to find all homogeneous solutions using the Frobenius method and the method of reduction of order. In Section \ref{sec:gensols} we develop an optimized strategy for finding the general solutions of coupled Feynman integrals. Lastly, we slightly improve the integration strategy of Ref. \cite{Frellesvig:2019byn} in Section \ref{sec:predivision}, by deriving explicit formulas for the center points of neighbouring line segments.

\subsection{Differential equations order-by-order in $\epsilon$}
\label{sec:deqnsordbyordeps}
Suppose that we have a line segment described by the path $\gamma(x)=(\gamma_{s_1}(x), \gamma_{s_2}(x), \ldots,\gamma_{s_{|S|}}(x))$, where $s_1,s_2,\ldots \in S$ denote the kinematic invariants and internal masses, and where $x$ is the line parameter. We may then write:
\begin{align}
    \label{eq:diffexpl}
    \partial_x \vec{f}(x,\epsilon) = \mathbf{A}_x(x,\epsilon) \vec{f}(x,\epsilon)\,,\quad \mathbf{A}_x = \sum_{s\in S}\mathbf{A}_s(\gamma(x)) \,\frac{\partial \gamma_s(x)}{\partial x}\,.
\end{align}
Let us expand the partial derivative matrix in terms of the dimensional regulator:
\begin{align}
    \label{eq:Axexp}
    \mathbf{A}_x(x,\epsilon) = \sum_{k = 0}^\infty \mathbf{A}_x^{(k)}(x) \epsilon^k\,.
\end{align}
We have assumed that there are no poles of the form $1/\epsilon^k$ for $k\geq 1$. Such poles in $\epsilon$ may typically be removed by rescaling the basis integrals with overall powers of $\epsilon$. We sum up to infinity in order to account for terms of the type $1/P(\epsilon)$, where $P(\epsilon)$ denotes a polynomial in $\epsilon$ with $P(0)\neq 0$. In general, it is also convenient to rescale the basis integrals by $\epsilon$-dependent factors that remove any terms of the form $1/P(\epsilon)$, whenever possible, so that for some positive integer $K$, it holds that $\mathbf{A}_x^{(k)} = 0$ for all $k>K$. This will speed up the computation of the series expansions of the master integrals at higher orders in $\epsilon$. We will assume the basis integrals are finite, which can be achieved by normalizing them with an overall power of $\epsilon$, and we will write their $\epsilon$ expansion as:
\begin{align}
    \label{eq:flaur}
    \vec{f}(x,\epsilon) = \sum_{k = 0}^\infty \vec{f}^{(k)}(x) \epsilon^k\,.
\end{align}
For brevity, we will drop the dependence on $x$ in the notation in the following. Plugging Eqns. (\ref{eq:Axexp}) and (\ref{eq:flaur}) into Eq. (\ref{eq:diffexpl}), and collecting terms order-by-order in $\epsilon$, we obtain:
\begin{align}
    \label{eq:startingpointdeqnsolve}
    \partial_x \vec{f}^{(k)} &= \mathbf{A}_x^{(0)}\vec{f}^{(k)} + \sum_{j=0}^{k-1} \mathbf{A}_x^{(k-j)} f^{(j)}\,.
\end{align}
It is clear that the matrix $\mathbf{A}_x^{(0)}$ plays a special role, as it multiplies the homogeneous component of the differential equations. Note that for a canonical basis $\mathbf{A}_x^{(0)} = 0$. In the following sections, we will solve Eq. (\ref{eq:startingpointdeqnsolve}) by considering sets of coupled integrals. Roughly spoken, we consider integrals to be coupled when their derivatives depend on each other at leading order in $\epsilon$, i.e. in the part that is expressed by the $\mathbf{A}_x^{(0)}$ matrix. We will make the definition of coupled integrals rigorous in Section \ref{sec:intorder}. 

Let $\{f_{\sigma_1},\ldots,f_{\sigma_p}\}$ be a set of coupled integrals, where $\Sigma = \{\sigma_1,\ldots,\sigma_p\}$ labels a subset of the master integrals. For convenience we will introduce the notation $f_{\sigma_1} \rightarrow g_1, f_{\sigma_2} \rightarrow g_2,$ and so on, and let $\vec{g} = (g_1,\ldots,g_p)$. We are then interested in the differential equations
\begin{align}
    \label{eq:coupledset}
    \partial_x \vec{g}^{(k)} = \mathbf{M} \vec{g}^{(k)} + \vec{b}^{(k)}\,,
\end{align}
where we have explicitly that:
\begin{align}
    \label{eq:coupledsetdecomposed}
    \mathbf{M}_{ij} =  (\mathbf{A}^{(0)}_x)_{\sigma_i,\sigma_j}\,, \quad \vec{b}_{i}^{(k)}=\sum_{j \notin \Sigma}\left[\left(\mathbf{A}_{x}^{(0)}\right)_{\sigma_{i} j} f_{j}^{(k)}+\sum_{l=0}^{k-1}\left(\mathbf{A}_{x}^{(k-l)}\right)_{\sigma_{i} j} f_{j}^{(l)}\right]
\end{align}
In the following sections we will discuss in detail how to solve Eq. (\ref{eq:coupledset}) as a series expansion around the origin. As a final remark, we will assume that the matrix $\mathbf{M}$ does not contain functions other than rational functions and square roots of irreducible polynomials. This also means that the basis of master integrals that we choose should not contain prefactors other than rational functions and square roots.

\subsection{Deriving an integration sequence}
\label{sec:intorder}
The first task in solving the differential equations is to determine an integration sequence. We should start by integrating the leading order in $\epsilon$ of the integrals, and move up one order in $\epsilon$ at a time, since the derivatives of the higher order terms contain contributions of the lower order terms (see Eq. (\ref{eq:startingpointdeqnsolve}).) Furthermore, for any given integral, its subsectors should be integrated first, since derivatives of subsectors never evaluate to terms containing integrals in higher sectors. Next, we show how to read off a suitable integration sequence directly from the partial derivative matrices, which can be done using basic graph theory.

First we define a new matrix $\mathbf{C}$, which is of the same size as $\mathbf{A}_x^{(0)}$ (i.e. $k\times k$ where $k$ is the number of master integrals), and which will be interpreted as the adjacency matrix of a directed graph $G$. We define $\mathbf{C}$ such that its elements $\mathbf{C}_{ij}$ are equal to one if $(\mathbf{A}_x^{(0)})_{ji}$ is nonzero, and zero otherwise. That way, the vertices of the directed graph $G$ are the basis integrals, and $G$ has an edge $j \rightarrow i$ for all nonzero $(\mathbf{A}_x^{(0)})_{ij}$. Next, consider the strongly connected components of $G$. Each strongly connected component is a set of vertices for which there is a directed path between every pair of vertices. Note that every vertex is connected to itself by the trivial path. By repeatedly differentiating an integral in a strongly connected component, one will eventually obtain a contribution from any other integral in the strongly connected component. We will call such integrals coupled, and their differential equations have to be solved simultaneously.

Next, consider the condensation $\tilde{G}$ of the graph $G$. This is the graph whose vertices are the strongly connected components of $G$, and which has an edge between components $c_1$ and $c_2$ if there is at least one directed edge in $G$ between a vertex of $c_1$ and a vertex of $c_2$. An integration sequence is then found by topologically sorting the vertices of $\tilde{G}$, meaning that a vertex $c_i$ comes before $c_j$ if there is a directed path from $c_i$ to $c_j$. For example, suppose we have three master integrals, and find the set $\{\{3\},\{1,2\}\}$ after sorting. This indicates that we should first integrate the third integral, and then integrate together the coupled integrals one and two. Note that in general topological sorting does not lead to a unique integration sequence, but we are free to pick any integration sequence that is compatible with the topological ordering. Lastly, we remark that we should (re-)derive an integration sequence for each path $\gamma(x)$. This is because sometimes integrals are coupled when transported along certain directions, but not along others. Luckily, deriving an integration sequence is very fast in the above approach.

\subsection{Homogeneous solutions and the Frobenius method}
\label{sec:homogeneoussolutions}
In the following section we discuss how to solve the homogeneous component of the differential equations of a set of coupled integrals as a series expansion around the origin of the line segment. We adopt the notation of Eq. (\ref{eq:coupledset}), but we will drop the superscripts, since the homogeneous differential equations are the same at each order in $\epsilon$. Thus, we are interested in solving differential equations of the form:
\begin{align}
    \label{eq:eqhom}
    \partial_x \vec{g} = \mathbf{M} \vec{g}\,,
\end{align}
for a vector of integrals $\vec{g} = (g_1,\ldots,g_p)$. For simplicity, we will use the notation $\partial = \partial_{x}$. Furthermore, we will let $g^{(j)} \equiv \partial^j \vec{g}$. Note that the superscript now does not refer to the order in $\epsilon$, which was the case in Section \ref{sec:deqnsordbyordeps}. We define a set of matrices $\mathbf{M}^{(j)}$ by:
\begin{equation}
    \label{eq:defAsups}
    \vec{g}^{(j)} \equiv \mathbf{M}^{(j)} \vec{g}\,.
\end{equation}
We can obtain these matrices by the recursion relation:
\begin{align}
\label{eq:Mmatrixrecursion}
\mathbf{M}^{(0)} = \mathbb{1}\,,&& \mathbf{M}^{(j)} = \partial \mathbf{M}^{(j-1)} + \mathbf{M}^{(j-1)} \mathbf{M}^{(1)}\quad\text{for all }j\geq 1\,.
\end{align}
Since we are interesting in finding series solutions, we expand $\mathbf{M}$ around the point $x=0$ up to a given order, and we compute $\mathbf{M}^{(j)}$ in terms of series expansions as well. Note that upon series expanding square roots, we have to take care that we choose the correct analytic branch of the square root. This is discussed in more detail in Section \ref{sec:analyticcontinuation}.

Next, consider the $(p \times p)$-matrix $\tilde{\mathbf{M}}$ whose rows are given by the top rows of the matrices $\mathbf{M}^{(j)}$. In particular, we have: $\tilde{\mathbf{M}}_{ij} = \mathbf{M}^{(i-1)}_{1j}$. Furthermore, consider the vector $\vec{g}^\partial = (g_1,\partial g_1, \ldots, \partial^{p-1} g_1)$. Then it holds that:
\begin{align}
    \label{eq:gpartvec}
    \vec{g}^\partial = \tilde{\mathbf{M}} \vec{g}\,.
\end{align}
If $\tilde{\mathbf{M}}$ is invertible, we may write:
\begin{align}
    \label{eq:relatepartials}
    \vec{g} = \tilde{\mathbf{M}}^{-1} \vec{g}^\partial \,.
\end{align}
For generic configurations of the kinematic invariants and internal masses, $\tilde{\mathbf{M}}$ is invertible. If the master integrals are instead integrated along line segments that lie on degenerate configurations of the kinematic invariants and internal masses, it may happen that there are relations between the master integrals along the line. In such cases, $\tilde{\mathbf{M}}$ might not be invertible. %This situation arises for example when the unequal mass banana graph families are evaluated on a line where the masses are equal. 
We will discuss a solution strategy for the case where $\tilde{\mathbf{M}}$ is singular at the end of this section, and we assume for now that $\tilde{\mathbf{M}}$ is invertible. Note that one way of avoiding the situation is to use a set of differential equations where the additional relations between the master integrals have been plugged in explicitly.

We are interested in finding and solving a $p$-th order differential equation for $g_1$. In particular, we seek a vector $\vec{c} = (c_{0}, \ldots, c_p)$, such that:
\begin{align}
    \label{eq:nthorder}
    \sum_{j=0}^p c_{j} g_1^{(j)} = 0\,.
\end{align}
Note that the elements of $\vec{c}$ depend on $x$. Consider the $((p+1)\times p)-$matrix $\tilde{\mathbf{M}}_+$, which is again defined by $(\tilde{\mathbf{M}}_+)_{ij} = \mathbf{M}^{(i-1)}_{1j}$. There is a unique vector $c^\intercal$ in the left null-space of $\tilde{\mathbf{M}}_+$, up to normalization, since we assumed that $\tilde{\mathbf{M}}$ is invertible. Next, define the vector $\vec{g}_+^\partial = (g_1,\partial g_1, \ldots, \partial^{p} g_1)$. We then obtain the desired differential equation in the following way:
\begin{align}
    c^\intercal \vec{g}^\partial_+ = c^\intercal\tilde{\mathbf{M}}_+\vec{g} = 0
\end{align}
We will normalize $c^\intercal$ such that $c_p=1$, i.e. the coefficient of the highest derivative is set to one. Next, we will discuss how to solve Eq. (\ref{eq:nthorder}) using a simple formulation of the Frobenius method. 

The Frobenius method is a general method for solving a homogeneous ordinary differential equation around a regular point $x = 0$, in terms of series expansions. The main idea relies on taking an ansatz for the solution in terms of a series of the form:
\begin{align}
    \label{eq:seriesansatz}
    g_1(x) = x^r s(x)\,,\quad s(x) = \sum_{m=0}^\infty s_{m} x^m\,
\end{align}
for some rational number $r$. We may series expand the coefficients of the differential equations, plug Eq. (\ref{eq:seriesansatz}) into Eq. (\ref{eq:nthorder}), and collect terms based on powers of $x$. We then obtain a set of equations for the coefficients $s_{m}$. At leading order in $x$, the equation is a non-trivial polynomial equation for $r$, which is called the indicial equation. The indicial equation will in general have multiple solutions. It turns out that if we take the largest solution for $r$, we may (recursively) solve for all $s_m$ with $m\geq 1$, by considering the equations defined by the remaining orders of $x$. The value of $s_0$ is a free parameter, and we may put it to one. The reason for picking the largest root of the indicial equation is to ensure that the recursion for $s_m$ does not break down. This can be seen if one works out the recursion symbolically, but we will not do that here (see e.g. Ref. \cite{Coddington} for a more detailed review of the Frobenius method.)

Thus, the Frobenius method yields at least one series solution to the differential equations. Next, we discuss how to find the $p-1$ remaining independent series solutions, using the well-known method of reduction of order. Let $D = \sum_{i=0}^p c_i \partial^i$ be the differential operator associated with Eq. (\ref{eq:nthorder}), and assume that $h$ is the solution from the Frobenius method, which satisfies $D h = 0$. Next, consider a multiplicative ansatz of the form $h\mu$, where $\mu = \int\nu$, which satisfies $D(h\mu)=0$. We then have explicitly:
\begin{align}
    \label{eq:multans}
    0 = D \left(h  \mu\right) = \sum_{j=0}^p c_j \partial^j \left(h  \mu\right) = \sum_{j=0}^p \sum_{n=0}^j c_j  {j\choose n} (\partial^{j-n} h) (\partial^{n} \mu)\,.
\end{align}
Note that the coefficient of $\mu = \partial^0\mu$, in the above equation, is simply given by:
\begin{align}
    \sum_{j=0}^p c_j \partial^{p} h = D h = 0\,.
\end{align}
Thus, Eq. (\ref{eq:multans}) is a $p$-th order differential equation for $\mu$ with no $\partial^0 \mu$-coefficient. Therefore, it defines a $(p-1)$-th order differential equation for $\nu$. We may describe this equation by a new differential operator $D' = \sum_{i=0}^{p-1} c_{i}' \partial^i$, for which we may again find one solution using the Frobenius method. It is clear that we may take another multiplicative ansatz, and iterate until we obtain a trivial differential equation. A possible recursive implementation in Mathematica looks as follows:
\begin{minted}[linenos=false]{text}
    FrobeniusSolutions[DEq_] := Block[{Sols = {}, DEq2},
        AppendTo[Sols, FrobeniusSolution[DEq]];
        
        If[DEqnOrder[DEq] > 1,  
            DEq2 = Dprime[DEq, Sols[[-1]]];
            Sols = Join[Sols, (Sols[[-1]] * Integrate[#, x])& /@ 
                    FrobeniusSolutions[DEq2]];
        ];
        
        Return[Sols]
    ];
\end{minted}
In the above example, the function \texttt{DEqnOrder[DEq\_]} represents a function that returns the order of the differential equation \texttt{DEq}. The function \texttt{FrobeniusSolution[DEq\_]} represents a function that returns a solution to the differential equation from the series ansatz in Eq. (\ref{eq:seriesansatz}), and lastly, the function \texttt{ReduceD[DEq\_, h\_]} represents a function that returns a lower order differential equation from the solution \texttt{h} given in the second argument. The series solutions which are obtained will contain terms of the type:
\begin{align}  
    \label{eq:serterms}
    \lambda^{i} \log (\lambda)^{j}\,,
\end{align}
where $i$ is a rational number, and $j$ is a non-negative integer. Such terms may be integrated in terms of combinations of terms of the same form, by repeatedly using an integration-by-parts identity to reduce the power of the logarithm down to zero. Within DiffExp, the integration of terms of the form of Eq. (\ref{eq:serterms}) is implemented using a list of replacement rules, which is faster than using the Mathematica function \texttt{Integrate[...]}, like in the above example.

We now have a way of obtaining $p$ independent solutions, which we will denote by $h_1,\ldots,h_p$ in the following. Next, consider the Wronskian matrix:
\begin{equation}
\label{eq:wronsk}
\mathbf{W}=\left|\begin{array}{ccc}
h_{1} & \cdots & h_{p} \\
\partial h_{1} & \cdots & \partial h_{p} \\
\vdots & \ddots & \vdots \\
\partial^{p-1} h_{1} & \cdots & \partial^{p-1} h_{p}
\end{array}\right|
\end{equation}
A matrix of solutions $\mathbf{F}$ to the homogeneous differential equation in Eq. (\ref{eq:eqhom}) is found by putting the Wronskian at the place of $\vec{g}^\partial$ in Eq. (\ref{eq:relatepartials}), which leads to:
\begin{align}
    \label{eq:Fmatdefinition}
    \mathbf{F} = \tilde{\mathbf{M}}^{-1} \mathbf{W}\,,\quad \partial \mathbf{F} = \mathbf{M} \mathbf{F}\,.
\end{align}
We may multiply the columns of $\mathbf{F}$ by free parameters, and sum over them, to obtain a general vector solution to Eq. (\ref{eq:eqhom}).
\subsection{General solutions}
\label{sec:gensols}
In the previous subsection, we showed how to solve homogeneous differential equations of the form of Eq. (\ref{eq:eqhom}). Next, we describe how to obtain the general solution to a system of differential equations of the type:
\begin{align}
    \label{eq:inhomdeqn}
    \partial_x \vec{g} = \mathbf{M} \vec{g} + \vec{b}\,,
\end{align}
which will allow us to solve Eq. (\ref{eq:coupledset}) in particular. First, consider the matrix:
\begin{align}
    \mathbf{B} = \frac{1}{p}(\vec{b},\ldots,\vec{b})\,,
\end{align}
where $\vec{b} = (b_{1},\ldots,b_{p})$ is a vector of size $p$. Next, consider the matrix $\mathbf{G} = \mathbf{F}\mathbf{H}$, which satisfies:
\begin{align}
    \partial \mathbf{G} = \mathbf{M}\mathbf{G} + \mathbf{B}\,,
\end{align}
where $\mathbf{F}$ is given in Eq. (\ref{eq:Fmatdefinition}), and where $\mathbf{H}$ will be determined next. We then have that:
\begin{align}
    \mathbf{F}\partial \mathbf{H} = \mathbf{B}\,\Rightarrow\, \mathbf{H} = \int \mathbf{F}^{-1}\mathbf{B} + \mathbf{E}\,,
\end{align}
where $\mathbf{E}$ is any constant matrix. We let $\mathbf{E}$ be a diagonal matrix of the form $\mathbf{E} = \text{diag}(e_1,\ldots,e_p)$, where the constants $e_j$ are to be fixed from boundary conditions. The general vector solution to Eq. (\ref{eq:inhomdeqn}) is then given by:
\begin{align}
    \label{eq:ggeneral}
    \vec{g} = \sum_{k=1}^p \vec{G_k}\,,\quad \mathbf{G} = \mathbf{F}\left(\int \mathbf{F}^{-1}\mathbf{B} + \mathbf{E}\right)\,,
\end{align}
where $\vec{G_k}$ denotes the $k$-th column of $\mathbf{G}$. In principle, this concludes the task of solving the differential equations. Let us discuss some optimizations to computing Eq. (\ref{eq:ggeneral}). Note that the definition of $\mathbf{F}$ relies on the inverse matrix $\tilde{\mathbf{M}}^{-1}$, while the definition of $\mathbf{G}$ also relies on the inverse matrix $\mathbf{F}^{-1} = \mathbf{W}^{-1} \tilde{\mathbf{M}}$. Since our matrix elements contain series expansions, it can be computationally expensive to compute these inverses when $p$ is large. Let us first consider the inverse of $\tilde{\mathbf{M}}$. Note that the entries of $\tilde{\mathbf{M}}$ contain series expansions without logarithms, since there were no integrations involved in computing $\tilde{\mathbf{M}}$. This makes the computation of the inverse of $\tilde{\mathbf{M}}$ relatively straightforward. In the current version of DiffExp, the Mathematica function \texttt{Inverse[...]} is used, with the option \texttt{Method} set to \texttt{"DivisionFreeRowReduction"}.

Next, let us consider the Wronskian matrix $\mathbf{W}$. Its entries contain series expansions which may contain logarithmic terms of the form $\log(x)$, and we find in this case that Mathematica has trouble to explicitly compute the inverse matrix, or an associated linear system, for high orders of $p$. We remedied this problem in a manner which we discuss next. First note that the Wronskian matrix satisfies a differential equation of the form:
\begin{equation}
    \partial \mathbf{W}=\mathbf{N}\mathbf{W}\,,\quad \mathbf{N} = \left(\begin{array}{cccccc}
0 & 1 & 0 & \cdots & 0 & 0 \\
0 & 0 & 1 & \cdots & 0 & 0 \\
\vdots & \vdots & \vdots & \ddots & \vdots & \vdots \\
0 & 0 & 0 & \cdots & 0 & 1 \\
-c_0 & -c_{1} & -c_{2} & \cdots & -c_{p-2} & -c_{p-1}
\end{array}\right)\,.
\end{equation}
Furthermore, we have:
\begin{equation}
    0 = \partial (\mathbf{W} \mathbf{W}^{-1}) = (\partial \mathbf{W} ) \mathbf{W}^{-1} + \mathbf{W} \partial \mathbf{W}^{-1} = \mathbf{N} + \mathbf{W} \partial \mathbf{W}^{-1}\,.
\end{equation}
Therefore, we have:
\begin{align}
    \label{eq:deqnwronskinv}
    \partial (\mathbf{W}^{-1})^\intercal = -\mathbf{N}^\intercal (\mathbf{W}^{-1})^\intercal\,.
\end{align}
We may solve this differential equation for $\mathbf{W}^{-1}$ using the Frobenius method following the steps outlined in Section \ref{sec:homogeneoussolutions}. After solving Eq. (\ref{eq:deqnwronskinv}) this way, we obtain a matrix, let us call it $\mathbf{X}^{-1}$, that is not quite the inverse of $\mathbf{W}$, but which satisfies the condition $\partial (\mathbf{W}\mathbf{X}) = \mathbb{1}$. Therefore, $\mathbf{W}\mathbf{X}$ is a constant matrix, which we will call $\mathbf{Z}$. We may easily invert $\mathbf{Z}$, and we can then obtain the inverse of the Wronskian matrix as $\mathbf{W}^{-1} = \mathbf{X}\mathbf{Z}^{-1}$. We use this approach to calculate $\mathbf{W}^{-1}$ in DiffExp when the Wronskian contains logarithms. Otherwise, we invert $\mathbf{W}$ directly, which we find to be faster in that case.

Note that we only compute $\mathbf{F}$ and $\mathbf{F}^{-1}$ once for each set of coupled integrals on a given line segment. To find the solutions of the coupled integrals at a given order in $\epsilon$, we then compute the appropriate $\mathbf{B}$-matrix, and use Eq. (\ref{eq:ggeneral}). Lastly, we remark that the above integration strategy is essentially equivalent to the method of variation of parameters, which we discuss next in Section \ref{sec:variationofparameters}. However, we found (by considering a number of examples) that the above way of computing the solutions is a bit more efficient in practice.

\subsection{Solutions along degenerate lines}
\label{sec:variationofparameters}
The integration strategy discussed in the previous section relies on the property that $\tilde{\mathbf{M}}$ is invertible, which is the case along generic contours where all master integrals are independent. We have also implemented a more direct version of the method of variation of parameters. We find that this method typically performs slower than the one discussed in Section \ref{sec:gensols}. However, we have found it more straightforward to generalize this method to the case where $\tilde{\mathbf{M}}$ is not invertible. We discuss the method next.

Consider the differential equations in Eq. (\ref{eq:inhomdeqn}), repeated here for clarity:
\begin{align}
    \label{eq:inhomdeqn2}
    \partial_x \vec{g} = \mathbf{M} \vec{g} + \vec{b}\,.
\end{align}
Next, define an analogue of Eq. (\ref{eq:defAsups}) by introducing vectors $\vec{\tilde{b}}^{(j)}$, so that:
\begin{equation}
    \vec{g}^{(j)} = \mathbf{M}^{(j)} \vec{g} + \vec{\tilde{b}}^{(j)}\,.
\end{equation}
We then have that:
\begin{align}
    \vec{\tilde{b}}^{(0)} = 0 \,,&& \vec{\tilde{b}}^{(j)} = \partial \vec{\tilde{b}}^{(j-1)} + \mathbf{M}^{(j-1)} \vec{b}\quad\text{for all }j\geq 1\,.
\end{align}
Next, we seek to find a higher order differential equation for each integral in $\vec{g}$. Consider the set of $((p+1)\times p)-$matrices $\tilde{\mathbf{M}}_{q,+}$, and vectors $\vec{\tsup[2]{b}}_{q,+}$ of length $(p+1)$, for $q = 1,\ldots, p$, defined by:
\begin{align}
    (\tilde{\mathbf{M}}_{q,+})_{ij} \equiv \mathbf{M}_{qj}^{(i-1)}\,,\quad (\vec{\tsup[2]{b}}_{q,+})_i \equiv \vec{\tilde{b}}_q^{(i-1)}\,.
\end{align}
Furthermore, let $\vec{g}_{q,+}^\partial = (g_q,\partial g_q, \ldots, \partial^{p} g_q)$. Then we have:
\begin{align}
    \vec{g}_{q,+}^\partial = \tilde{\mathbf{M}}_{q,+}\vec{g} + \vec{\tsup[2]{b}}_{q,+}\,.
\end{align}
Next, let $c_q^\intercal$ denote the vector (up to normalization) in the left null-space of $\mathbf{M}_{q,+}$ with the most trailing zeros (i.e. which gives the lowest order differential equation for integral $q$.) Then we obtain the following differential equations:
\begin{align}
    \label{eq:higherorderdeqnfull}
    c_{q}^\intercal \vec{g}_{q,+}^\partial = \underbrace{c_{q}^\intercal\tilde{\mathbf{M}}_{q,+}\vec{g}}_{=0} + c_{q}^\intercal \vec{\tsup[2]{b}}_{q,+} = 0\,,\quad \text{ for all }q=1,\ldots,p\,.
\end{align}
Let us consider the differential equation for the integral $g_q$. We will denote the order of the differential equation by $v_q$. We have:
\begin{align}
    \label{eq:higherorderdeqnfullexpl}
    \sum_{j=0}^{v_q} c_{q,j} \partial^j g_q + \tsup[3]{b}_q = 0\,,
\end{align}
where $\tsup[3]{b}_q\equiv -c_{q}^\intercal \vec{\tsup[2]{b}}_{q,+}$. We choose the normalization $c_{q,v_q} = 1$. We may obtain the homogeneous solutions to Eq. (\ref{eq:higherorderdeqnfull}) using the Frobenius method, as described in Section \ref{sec:homogeneoussolutions}. Denote these by $h_{q,1},\ldots, h_{q,v_q}$. The method of variation of parameters tells us that the general solution to Eq. (\ref{eq:higherorderdeqnfullexpl}) can then be written as:
\begin{align}
    \label{eq:gqvarofpar}
    g_q = \sum_{j=1}^{v_q} h_{q,j} \left(e_{q,j} + \int \frac{\mathbf{W}_{(q,j)}}{\mathbf{W}_{(q)}}\,dx\right)\,,
\end{align}
where the constants $e_{q,j}$ are are to be determined from boundary conditions, where $\mathbf{W}_{(q)}$ is the Wronskian determinant of the homogeneous solutions $h_{q,1},\ldots, h_{q,v_q}$, and where $\mathbf{W}_{(q,j)}$ is the determinant of the Wronskian matrix with the $j$-th column replaced by the vector $(0,\ldots,0,\tsup[3]{b}_q)$. 

Next, we distinguish two cases. In the first case, we consider that $\tilde{\mathbf{M}}_q$ is invertible for some $q$, where $\tilde{\mathbf{M}}_q$ is the $(p\times p)$-matrix obtained by removing the last row of $\tilde{\mathbf{M}}_{q,+}$. We may then compute Eq. (\ref{eq:gqvarofpar}) for the given $q$, and use that:
\begin{equation}
    \vec{g} = \tilde{\mathbf{M}}_{q}^{-1}\left(\vec{g}_{q}^\partial - \vec{\tsup[2]{b}}_{q}\right)\,,
\end{equation}
in order to find the solutions of the other basis integrals. Here we used the notation $\vec{g}_{q}^{\partial}$ and $\vec{\tsup[2]{b}}_{q}$ to denote the vectors $\vec{g}_{q,+}^{\partial}$ and $\vec{\tsup[2]{b}}_{q,+}$ with the last entry removed. In the second case where $\tilde{\mathbf{M}}$ is not invertible, we can use Eq. (\ref{eq:gqvarofpar}) to compute all the coupled integrals. Because the integrals are related through Eq. (\ref{eq:inhomdeqn2}), there are relations between the constants $e_{q,j}$ for different $q$. These are fixed by plugging the solutions of Eq. (\ref{eq:gqvarofpar}) into Eq. (\ref{eq:inhomdeqn2}) and eliminating the redundant constants by reducing the resulting linear system. 

Note that we only have to compute the matrix $\mathbf{\tilde{W}}_{(q)}$ and the matrices $\mathbf{M}^{(j)}$ once for a given line segment. For each order in $\epsilon$, we then compute the corresponding terms $\tsup[3]{b}_q$ and determinants $\mathbf{W}_{(q,j)}$. The evaluation of the determinants can be computationally heavy when the number of integrals $p$ is large. The method of variation of parameters may be enabled in DiffExp by setting the option \texttt{IntegrationStrategy} to the value \texttt{"VOP"} in the configuration.

\subsection{Analytic continuation}
\label{sec:analyticcontinuation}
In the following section we discuss how to perform the analytic continuation of the series solutions to the differential equations. We also discuss some specific details related to DiffExp. 

If we solve the differential equations on a line segment that is centered on a threshold singularity, the series expansions may contain multivalued functions of the form $\log(x)$ and $\sqrt{x}$. Square roots may arise from a Frobenius ansatz of the form of Eq. (\ref{eq:seriesansatz}), when the maximal root $r$ of the indicial equation has denominator two. We are not aware of any Feynman integral family for which there are homogeneous solutions containing roots of degree higher than two. Square roots may also appear when the partial derivative matrices contain square roots. DiffExp is only able to analytically continue square roots and logarithms. Therefore, the user should give a set of differential equations that contains only rational functions and square roots.

We can specify the branch of the logarithms and square roots by adding an infinitesimally small imaginary part to the argument. For real $x$, we have:
\begin{align}
    \label{eq:analyticcontinuation}
    &\log(x+i\delta) = \log(x)\,, && \sqrt{x+i\delta} = \sqrt{x}\,, \nonumber\\
    &\log(x-i\delta) = \log(x) -2\pi i  \theta_m\,, && \sqrt{x-i\delta} = (\theta_p-\theta_m)\sqrt{x}\,,
\end{align}
where we let $\theta_p = \theta(x)$ and $\theta_m = \theta(-x)$ be Heaviside step functions. Within the Mathematica code, we don't have to work with terms of the form $i\delta, \theta_p$ and $\theta_m$ explicitly, but we can instead implement the above relations using replacement rules. For example, if we know the line parameter carries a small negative imaginary part, we can evaluate the series expansions at negative values of the line parameter by applying the following rules before evaluation:
\begin{align}
    \label{eq:analyticcontinuationreplacementrules}
    \log(x) \rightarrow \log(x) - 2\pi i\,, && \sqrt{x} \rightarrow -\sqrt{x}\,.
\end{align}
Internally DiffExp only uses replacement rules, but results that are provided to the user from the function \texttt{IntegrateSystem[...]} carry explicit factors of $\theta_p$ and $\theta_m$.

If we seek to obtain results in a given physical region, the imaginary part of the line parameter should be in correspondence with the Feynman prescription. We discuss next how this is handled in DiffExp. First, note that the Feynman prescription does not always provide a unique choice of signs of the Mandelstam variables. We saw in Section \ref{sec:analyticcontinuationremarks} that every variable $s_{(T_1,T_2)}$ should carry a positive imaginary part, and that the masses should carry a negative imaginary part. However, the quantities $s_{(T_1,T_2)}$ evaluate to sums of Mandelstam variables, and are related by momentum conservation. Therefore, the Feynman prescription may sometimes be ambiguous in terms of the Mandelstam variables.

Therefore, we have taken a more general approach to analytic continuation in DiffExp, where instead of assigning an imaginary part to each Mandelstam variable, the user provides a list of polynomials with an additional term $\pm i\delta$, such that the zeros of the polynomials parametrize either physical threshold singularities or the vanishing locus of square roots, and such that the $i\delta$'s determines the choice of branch. This is done with the configuration option \texttt{DeltaPrescriptions}. For example, a user may put:
\begin{minted}[linenos=false]{text}
    DeltaPrescriptions -> {4msq-s-Iδ, 4*msq*(p4sq - s - t) + s*t + Iδ}
\end{minted}
to fix the $i\delta$-prescription for a threshold singularity at $s=4m^2$, where $s$ and $-m^2$ carry $+i\delta$, and to tell DiffExp that the square root $\sqrt{4 m^{2}\left(p_{4}^{2}-s-t\right)+s t}$ should be interpreted as:
\begin{equation}
    \sqrt{4 m^{2}\left(p_{4}^{2}-s-t\right)+s t+i\delta}\,.
\end{equation}
To transfer the user-provided $i\delta$-prescriptions over to the line parameter $x$, we plug the line into the polynomials provided by the user, expand the resulting expressions in $x$, and take the leading terms in $x$. The polynomials for which the leading term is constant are discarded, as the current line segment is not centered on their corresponding singular surface. For the remaining polynomials, we check whether the leading term is proportional to $x$ (raised to power one.) If so, we can read off how to associate the $i\delta$-prescription with the line parameter. If the leading term is proportional to a different power of $x$, we are unable to transfer the Feynman prescription, and we should pick a different line segment.

Let us work out a simple example. Consider the square root $f(s) = \sqrt{4-s-i\delta}$. Its differential equation is given by:
\begin{align}
    \partial_s f(s) = -\frac{1}{2(4-s)} f(s)\,.
\end{align}
Let us consider the line $s = 4 - x$. We then have:
\begin{align}
    \label{eq:exlinediffeqns}
    \partial_x f(x) = \frac{1}{2 x}f(x)\,.
\end{align}
Solving the above differential equation leads to the general solution $f(x) = c_1 x^{1/2}$, where $c_1$ is to be fixed from boundary conditions. Furthermore, we have that $4-s(x) - i\delta = x - i\delta$. Hence, we see that $x$ carries the imaginary part $-i\delta$. Next, we use Eq. (\ref{eq:analyticcontinuation}) to update our general solution to the correct prescription, which gives:
\begin{align}
    f(x) = c_1 \sqrt{x-i\delta} = c_1 (\theta_p-\theta_m)\sqrt{x}\,.
\end{align}
Fixing $c_1$ at a boundary point gives $c_1 = 1$, which provides the correct answer.

We conclude this section with some additional remarks about the handling of square roots in the differential equations. Firstly, in the case where the square root lies on a physical threshold singularity, it is important that it is assigned the branch that agrees with the Feynman prescription. For example, if the basis contains a square root of the form $\sqrt{4m^2-s}$, and the Feynman prescription tells us that $s$ carries a positive imaginary part and $m^2$ carries a negative imaginary part, the corresponding square root should be interpreted as $\sqrt{4m^2-s-i\delta}$. The branch of the square roots which do not lie on a physical threshold singularity can be chosen freely. For those, it is most convenient to pick $+i\delta$, so that the square roots are given in the principal branch. All square roots for which the argument or $-1$ times the argument is not passed to the option \texttt{DeltaPrescriptions} by the user, will automatically be assigned the imaginary part $+i\delta$. 

Another point is that we should only work with square roots that contain irreducible arguments (over the real numbers).
If the arguments of the square roots are reducible the following problem could arise. Consider the square root $\sqrt{s(4-s)}$. We want to assign the square root some branch, for example $\sqrt{s(4-s)} \rightarrow \sqrt{s(4-s)+i\delta}$. Furthermore, suppose that the Feynman prescription dictates that $s$ carries a positive imaginary part. Along the line $s=x$, we can safely let $x\sim x+i\delta$, and this agrees with the choice of branch of $\sqrt{s(4-s)}$. However, along the line $s=4+x$, that would yield $\sqrt{s(x)(4-s(x))} = \sqrt{-x(4+x)} \sim \sqrt{-x(4+x)-i\delta}$. Therefore, we can not simultaneously satisfy the Feynman prescription and the fact that the square root is on the principal branch.

Lastly, note that upon series expanding the partial derivative matrices, we have to take care that the square roots are expanded in the correct branch. A simple way to do this is to take all square roots in the matrices which carry a $-i\delta$, and use the relation:
\begin{align}
    \sqrt{a(S)-i\delta} = -i\sqrt{-a(S)+i\delta}\,,
\end{align}
where $a(S)$ denotes an irreducible polynomial in the kinematic invariants and internal masses (which are denoted by the set $S$.) After using the above relation, we can use Mathematica's function \texttt{Series[...]} with the option \texttt{Assumptions $\rightarrow x>0$}, in order to obtain a series expansion that is valid for positive values of $x$. We can then evaluate the expansions at negative values of $x$, by using the replacement rules of Eq. (\ref{eq:analyticcontinuationreplacementrules}) whenever $x$ carries negative imaginary part.

\subsection{Precision and numerics}
\label{sec:precisionandnumerics}
In the following subsection we discuss two ways that we may increase the precision of the series expansions along a given line segment. First, we give a few remarks on the convergence of the expansions and the growth of the series coefficients.

\subsubsection{Convergence radius and growth of series coefficients}
\label{sec:precisionandconvergence}
Note that DiffExp is only designed to work with differential equations whose coefficients are composed of rational functions and square roots of rational functions. Suppose that we expand along a line parametrized by the line parameter $x$. The differential equations have singularities in the complex plane of $x$, at the positions of the poles of the rational functions, and the positions of the zeros of the arguments of the square roots. Let us denote these by the set $X_\text{sing} = \{x_1,\ldots,x_n\}$, and suppose that we are expanding around the origin $x=0$. Then, the radius of convergence of the series expansions is given by $r=\text{min}\{|x_i| \in X_\text{sing}\}$. Note that the Feynman integrals themselves typically possess a small subset of the singularities that are contained in the differential equations. Nonetheless, at intermediate stages of the calculations, our computations will be sensitive to the points in $X_\text{sing}$.

Within the DiffExp package, the coefficients of the series expansions are treated as inexact numbers which are valid up to the user-provided working precision. If $r$ is very small, we will typically find that the series coefficients of the expansions grow very fast. This is undesirable, as it may lead to loss of precision and/or numerical instabilities. For this reason, it is a good idea to rescale the line parameter in order to map the points in $X_\text{sing}$ away from the origin. In the upcoming sections, we will discuss two different segmentation strategies for the transportation of boundary conditions. In both strategies, the line segments which are returned are always chosen such that the radius of convergence $r$ satisfies $r\geq 1$. In some cases, the choice of $r=1$ may still lead to numerical instabilities. We find this to be the case for the unequal mass three-loop banana graph family, which is solved in Section \ref{sec:fourmassunequalbanana}. Therefore, DiffExp contains the additional option \texttt{RadiusOfConvergence}, which is equal to one by default. For values different than one, the line parameter will be rescaled so that for each segment $r$ is at least equal to the value of \texttt{RadiusOfConvergence}.

\subsubsection{Improving the precision: Möbius transformations}
\label{sec:mobiustransformations}
One way to improve the precision along a given line segment is to act with a specific Möbius transformation on the line parameter, which repositions the nearest singularities so that they are at an equal distance from the origin \cite{Frellesvig:2019byn}.

Suppose that we are interested in expanding around the origin of the line $\gamma(x)$. Furthermore, suppose that $X_{\text{sing}}=(x_1,\ldots,x_k) / \{0\}$ is a finite set of points at which the line $\gamma(x)$ crosses a singularity of the differential equations. We exclude the point zero from the set, as we are expanding at origin. Assume for now that all $x_j \in X_{\text{sing}}$ are real. We comment on the more general case later. Let $x_L < 0$ and $x_R > 0$ be the two points in $X_{\text{sing}}$ that are closest to the origin. If there is no $x_j\in X_{\text{sing}}$ such that $x_j < 0$, we let $x_L = -\infty$. Similarly, if there is no $x_j>0$ in $X_\text{sing}$, then we let $x_R = +\infty$. Now, consider a new line parameter $y$ defined by the Möbius transformation:
\begin{align}
    \label{eq:mobius}
    x(y) = \frac{2 y x_{L} x_{R}}{x_{L}-x_{R}+y\left(x_{L}+x_{R}\right)}\,,
\end{align}
such that the points $y=-1,0,1$ correspond to $x=x_L,0,x_R$ respectively. When $x_L = -\infty$, or $x_R=\infty$, we can take a limit of the Möbius transformation. Let $Y_\text{sing}=(y_1,\ldots,y_k) / \{0\}$ be the set of points such that $x(y_j) = x_j$. We then have that $|y_j| \geq 1$ for all $y_j$. Therefore, if we expand in the line parameter $y$, the resulting expansions converge in the range $y \in (-1,1)$, which corresponds to the range $(x_L, x_R)$ in the line parameter $x$. Had we expanded instead in the line parameter $x$, the expansions would have been valid in the smaller range $(-r,r)$, where $r=\text{min}(-x_L,x_R)$.

Let us illustrate this with a simple example. Consider the function:
\begin{align}
    \label{eq:fexample}
    f(x) = \frac{1}{1/10+x} - \frac{1}{1-x}\,,
\end{align}
which has poles at $x = -1/10$ and $x = 1$. We are interested in a series expansion at the point $x = 0$, which we denote by $S(f)(x)$. The first five orders are given by:
\begin{align}
    \label{eq:sfxser}
    S_5(f)(x) = 9-101 x+999 x^{2}-10001 x^{3}+99999 x^{4}-1000001 x^{5}+\mathcal{O}(x)^{6}\,.
\end{align}
It it clear that the series coefficients quickly grow in size, and that the radius of convergence of $f(x)$ is equal to $1/10$. Next, consider the line parameter $y$ defined by:
\begin{align}
    x(y) = -\frac{2 y}{9y-11}\,,
\end{align}
We may plug $x(y)$ into Eq. (\ref{eq:fexample}) or Eq. (\ref{eq:sfxser}), and obtain:
\begin{align}
    S_5(f)(y) = 9-\frac{202 y}{11}+18 y^{2}-\frac{202 y^{3}}{11}+18 y^{4}-\frac{202 y^{5}}{11}+\mathcal{O}\left(y^{6}\right)\,.
\end{align}
Notice that the series coefficients are now much better behaved. Furthermore, consider the point $x = 1/4$, which corresponds to $y=11/17$. Evaluating the series expansions up to order 15 gives:
\begin{align}
    S_{15}(f)(x) \approx -6.65\cdot 10^{6}\,, && S_{15}(f)(y) \approx 1.51\,, && f(x=1/4) = 32/21 \approx 1.52\,.
\end{align}
Clearly, the series in $x$ does not converge, while the series in $y$ does. Therefore, it is beneficial to use a Möbius transformation to remap the singularities.

Lastly, we comment on the case where some of the $x_j\in X_{\text{sing}}$ are complex numbers. In that case, we may consider instead the set $X_\text{sing}'$, which contains all the points $\text{Re}(x_j)$ for $x_j \in X_{\text{sing}}$ with $\text{Re}(x_j)\neq 0$, and which contains the points $\pm \text{Im}(x_j)$  where $x_j$ is the closest point to the origin satisfying $\text{Re}(x_j) = 0$. We may then proceed as before, with $X_{\text{sing}}$ replaced by $X_{\text{sing}}'$, and consider the line parameter $y$ of Eq. (\ref{eq:mobius}). In the complex case, it is not guaranteed anymore that expanding in $y$ is better than expanding in $x$. For example, there may be complex singularities with large imaginary parts, but real parts close to the origin, and it might not be optimal to map their real part to $-1$ or $+1$. One solution would be to increase $x_L$ and $x_R$ dynamically until one of the singularities in the complex plane of $y$ lands inside the unit disc, and such that $x_R - x_L$ is as large as possible, but we leave this for a future version of DiffExp.

\subsubsection{Improving the precision: Padé approximants}
If a function $f(x)$ admits a Taylor series at $x=0$, we may compute its Padé approximant of order $(n,m)$, which is a rational approximation to $f(x)$ of the form:
\begin{align}
    P_{n,m}(f)(x) = \frac{S_1(x)}{S_2(x)}\,,
\end{align}
where $S_1(x)$ and $S_2(x)$ are polynomials of degrees $n$ and $m$ respectively. The Padé approximant is uniquely defined by the property that its Taylor expansion matches the Taylor expansion of $f(x)$ up to order $\mathcal{O}(x^{n+m+1})$, and can be computed using standard algorithms. It is well-known in the field of applied mathematics that Padé approximants often yield a better approximation to a function than the Taylor series. (See e.g. Ref. \cite{bender2013advanced} for a more general overview of Padé approximants and series acceleration methods.) Furthermore, the Padé approximant is computed directly from the Taylor series of $f(x)$.

The definition of the Padé approximant can be extended to cover functions $f(x)$ which admit a Laurent expansion at $x=0$. In this case, we may multiply out the highest degree pole, compute a Padé approximant of the resulting Taylor series, and divide out the pole, to obtain a Padé approximant for $f(x)$. We can also extend the definition to power series with fractional powers. For example, suppose we have a series of the type:
\begin{align}
    \label{eq:laurfracseries}
    \sum_{j=-p}^k f_j x^{j/r} + \mathcal{O}(x^{(k+1)/r})\,,
\end{align}
for integers $k\geq -p$, and a positive integer $r \geq 1$. We may compute the Padé approximant of $\sum_{j=-p}^k f_j x^{j}$, and replace every power $x^j$ by $x^{j/k}$ afterwards. In this case the Padé approximant is not anymore the ratio of two polynomials, but of two power series with fractional powers. We will then let $P_{n,m}(f)(x)$ denote the Padé approximant where the powers of $x$ in the numerator are at most equal to $n$, and where the powers of $x$ in the denominator are at most equal to $m$. The Padé approximant is implemented in Mathematica, including for series with fractional powers, and can be called using the function \texttt{PadeApproximant[f[x], \{x, 0, \{n, m\}\}]}.

Note that series solutions of Feynman integrals may also contain powers of logarithms. To deal with these, we decompose the series expansions as:
\begin{align}
    \sum_{i=0}^q \log(x)^i \sum_{j=-p}^\infty f_{ij} x^{j/r}\,,
\end{align}
for a non-negative integer $q$, an integer $p$, and where $r = 1$ or $r = 2$. We then compute a Padé approximant for each power of the logarithm.

We may employ Padé approximants in our setup whenever we need to evaluate the series solutions of the Feynman integrals. For example, in order to compute boundary conditions for the next line segment, we compute the Padé approximant of each integral, and evaluate it at the next boundary point. Note that Padé approximants were also used in Ref. \cite{Frellesvig:2019byn} to improve the numerical precision. Lastly, note that DiffExp always computes the diagonal Padé approximant. In particular, for a series of the form of Eq. (\ref{eq:laurfracseries}), we let $n=\floor{\frac{k+p+1}{2}}$ and $m = \floor{\frac{k+p+1}{2}}$.

There are two possible caveats when using Padé approximants, which we discuss next. Firstly, note that the series solutions that are found by DiffExp have inexact numerical coefficients which are valid up to a certain number of digits. Typically, the accuracy of the coefficients of the Padé approximants is lower than the accuracy of the coefficients the original series. Therefore, when Padé approximants are enabled in DiffExp, the working precision should typically be increased too. The second caveat is that it can take some time to compute the Padé approximants of all the basis integrals, especially when the expansions contain half-integer powers and/or logarithms. Nonetheless, in the examples on which DiffExp was tested, we have almost always found Padé approximants to significantly decrease the computation time needed to obtain results at a given precision. However, to be safe, Padé approximants are disabled in DiffExp by default. They can be turned on by setting the option \texttt{UsePade} to True. 

\subsection{Line segmentation strategies}
In the following section we describe two strategies for transporting boundary conditions along a line, based on subdividing the line into multiple segments. First, we consider a "dynamic" segmentation strategy, in which we keep the error of the series expansions of the differential equations within a certain bound. In practice, bounding the error of the expansions of the differential equations also bounds the error of the series solutions to the differential equations (although not necessarily to the exact same extent.) Secondly, we describe a variation of the first strategy, which we call the "predivision" segmentation strategy. In that strategy, we subdivide the line into multiple segments, with the requirement that the expansions on each line segment are only evaluated at a fixed fraction of the distance to the nearest singularity of the differential equations. Similar integration strategies were considered in Refs. \cite{Francesco:2019yqt, Bonciani:2019jyb, Frellesvig:2019byn, Abreu:2020jxa}. Our implementation of the predivision strategy is closest to that of Ref. \cite{Frellesvig:2019byn}, where the strategy was considered with the use of Möbius transformations. Compared to that paper, we have slightly improved the matching of neighbouring line segments.

\subsubsection{Dynamic segmentation strategy}
\label{sec:dynseg}
Suppose we are transporting along a line $\gamma(x)$, from a point $x_{\text{start}}$ to a point $x_{\text{end}} > x_{\text{start}}$. Let $X_{\text{sing}} = (x_1, \ldots, x_k)$ denote the set of singularities of the differential equations in the complex plane of $x$, and assume that $x_{\text{end}} \notin X_{\text{sing}}$. Next, let
\begin{align}
    x(y_{\tilde{x}}) = \tilde{x} + r_{\tilde{x}} y_{\tilde{x}}
\end{align}
define a line parameter $y_{\tilde{x}}$, for each point $\tilde{x}$, such that $x(y_{\tilde{x}} = 0) = \tilde{x}$ and such that $x(y_{\tilde{x}} = \pm 1) = \tilde{x} \pm r_{\tilde{x}}$. The variable $r_{\tilde{x}}$ denotes the distance of $\tilde{x}$ to the nearest point in $X_{\text{sing}}$. By including this additional rescaling, the series expansions in $y_{\tilde{x}}$ behave better numerically, as discussed in Section \ref{sec:precisionandconvergence}, and converge within the interval $(-1,1)$. Next, consider some small number $\delta$, and define the interval
\begin{align}
    \label{eq:intytildex}
    I^\delta(y_{\tilde{x}}) &= [-y^\delta_{\tilde{x}} , y^\delta_{\tilde{x}}]\,,
\end{align}
where $y_{\tilde{x}}^\delta$ is the maximum real number so that:
\begin{align}
    \label{eq:expansionbound}
    \left|\mathbf{A}_{y_{\tilde{x}},ij}^{(k)}(y_{\tilde{x}})-S_{n} \left(\mathbf{A}_{y_{\tilde{x}},ij}^{(k)}\right)(y_{\tilde{x}})\right|<\delta\,, \quad \text{ for all }i,j,k,   \text{ and} -y^\delta_{\tilde{x}} < y_{\tilde{x}} < y^\delta_{\tilde{x}}\,,
\end{align}
and where $S_{n} \left(\mathbf{A}_{y_{\tilde{x}},ij}^{(k)}\right)$ denotes the series expansion of $\mathbf{A}_{y_{\tilde{x}},ij}^{(k)}$ in $y_{\tilde{x}}'$ up to order $\mathcal{O}(y_{\tilde{x}}^{n+1})$, where $n$ is the order at which the expansions are performed. In fact, within DiffExp, we usually take $n$ in Eq. (\ref{eq:expansionbound}) to be a few orders smaller than the order at which we perform the expansions, to be safe. The matrices $\mathbf{A}_{y_{\tilde{x}}}^{(k)}$ are the partial derivative matrices introduced in Eq. (\ref{eq:Axexp}), with respect to the line parameter $y_{\tilde{x}}$. In practice, it is hard to compute $y^\delta_{\tilde{x}}$ exactly, and so we instead use the estimate:
\begin{align}
    y^{\delta,\text{est}}_{\tilde{x}} = \min \left\{ \left(\frac{\delta}{\left|S^{(n)} \left(\mathbf{A}_{y_{\tilde{x}},ij}^{(k)}\right)\right|}\right)^{\frac{1}{n}}\,,\,  \text{ for all }i,j,k\right\}\,,
\end{align}
where $S^{(n)} \left(\mathbf{A}_{y_{\tilde{x}},ij}^{(k)}\right)$ denotes the coefficient of the $n$-th power $y_{\tilde{x}}^n$ in the series expansion of $\mathbf{A}_{y_{\tilde{x}},ij}^{(k)}$, and where we pick the real and positive $n$-th root. Lastly, let:
\begin{align}
    \label{eq:intervx}
    I^\delta(\tilde{x}) &= [\tilde{x}^{\delta,L}, \tilde{x}^{\delta,R}]\,,
\end{align}
where:
\begin{align}
    \tilde{x}^{\delta,L} = x(y_{\tilde{x}} = -y_{\tilde{x}}^\delta) && \tilde{x}^{\delta,R} = x(y_{\tilde{x}} = y_{\tilde{x}}^\delta)\,,
\end{align}
be the interval $I^\delta(y_{\tilde{x}})$ expressed in the line parameter $x$. Note that to compute these intervals, we have to perform the expansions of the partial derivative matrices in $y_{\tilde{x}}$, which can be time-consuming. The first step in our integration algorithm is to compute $I^\delta(x_i)$ for all real $x_i \in X_{\text{sing}}$, such that $x_\text{start} \leq x_i < x_\text{end}$.

Next, suppose that we are given boundary conditions $\vec{f}(x_{\text{bc}})$ at the point $x_{\text{bc}}$, and that we are given the point $x_\text{xp}$ on which to center the current line segment. Then we perform the following steps:
\begin{itemize}
    \item[1.] Expand the differential equations in $y_{x_{\text{xp}}}$ and find the corresponding series solutions. Fix boundary conditions at the point $y_{x_{\text{xp}},\text{bc}}$, where $x(y_{x_{\text{xp}}} = y_{x_{\text{xp}},\text{bc}}) = x_{\text{bc}}$.
    \item[2.] Determine $y^\delta_{x_{\text{xp}}}$ and $x_{\text{xp}}^{\delta,R}$. If $x_{\text{xp}}^{\delta,R} > x_{\text{end}}$, evaluate the series solutions at $y_{x_{\text{xp}},\text{end}}$, where $x(y_{x_{\text{xp}}} = y_{x_{\text{xp}},\text{end}}) = x_{\text{end}}$, and return the result. Otherwise, let $x_{\text{bc}}' = x_{\text{xp}}^{\delta,R}$, and evaluate the series solutions at $y^\delta_{x_{\text{xp}}}$, to obtain the next set of boundary conditions $\vec{f}(x_{\text{bc}}')$.
    \item[3.] If $x_{\text{bc}}' \in I^\delta(x_i)$ for some $x_i \in X_{\text{sing}}$, then let $x_{\text{xp}}' = x_i$. Otherwise, let $x_{\text{xp}}' = x_{\text{bc}}'$.
\end{itemize}
By iterating the above steps, starting with $x_{\text{xp}} = x_{\text{bc}} = x_{\text{start}}$, we may reach the endpoint $x_{\text{end}}$. Note that we may let $x_{\text{start}}\in X_\text{sing}$ if we give the set of boundary conditions $f(x_{\text{start}})$ as an asymptotic limit in the line parameter $y_{x_{\text{start}}}$.

Lastly, we discuss how we may incorporate Möbius transformations in the above setup, in the spirit of Section \ref{sec:mobiustransformations}. In this case we have to be careful with the presence of complex singularities. We may deal with complex singularities by defining a new set of points $X_{\text{sing}}'$ which consists of particular projections of the singularities in $X_{\text{sing}}$ onto the real axis. In particular, for each $x_i \in X_{\text{sing}}$, consider the set $X_{\text{proj}}(x_i)$, such that:
\begin{itemize}
    \item $\operatorname{Re}\left(x_{i}\right) \in X_{\text{proj}}\left(x_{i}\right)$,
    \item $\operatorname{Re}\left(x_{i}\right)-\operatorname{Im}\left(x_{i}\right) \in X_{\text{proj}}\left(x_{i}\right)$ if $!\left( \exists \,x_{j} \in X_{\text {sing }} | \operatorname{Re}\left(x_{i}\right)-\operatorname{Im}\left(x_{i}\right)<\operatorname{Re}\left(x_{j}\right)<\operatorname{Re}\left(x_{i}\right)\right)$,
    \item $\operatorname{Re}\left(x_{i}\right)+\operatorname{Im}\left(x_{i}\right) \in X_{\text{proj}}\left(x_{i}\right)$ if $! \left( \exists\,x_{j} \in X_{\text {sing }} | \operatorname{Re}\left(x_{i}\right)<\operatorname{Re}\left(x_{j}\right)<\operatorname{Re}\left(x_{i}\right)+\operatorname{Im}\left(x_{i}\right)\right)$\,.
\end{itemize}
Then we let $X_{\text{sing}}' = \cup_{x_i \in X_{\text{sing}}} X_\text{proj}(x_i) = \{x_1', \ldots, x_{k}'\}$. Next, we choose line parameters of the following form:
\begin{align}
    \label{eq:lineparsmobius}
    x(y_{\tilde{x}}) = \frac{y_{\tilde{x}}\left(2 \tilde{x}_{L} \tilde{x}_{R}-\tilde{x} \tilde{x}_{L}-\tilde{x} \tilde{x}_{R}\right)+\tilde{x} \tilde{x}_{L}-\tilde{x} \tilde{x}_{R}}{y_{\tilde{x}}\left(\tilde{x}_{L}+\tilde{x}_{R}-2 \tilde{x}\right)+\tilde{x}_{L}-\tilde{x}_{R}}\,,
\end{align}
where $\tilde{x}_L$ is the nearest point in $X_\text{sing}'$ that is on the left of $\tilde{x}$, and similarly for $\tilde{x}_R$. If there is no singularity on the left, we choose $\tilde{x}_L = -\infty$, and if there is no singularity on the right, we choose $\tilde{x}_R = +\infty$. We may now proceed with the same three integration steps as before, using the line parameters of Eq. (\ref{eq:lineparsmobius}), and replacing $X_\text{sing}$ by $X_{\text{sing}}'$ in the third step.

\subsubsection{Predivision segmentation strategy}
\label{sec:predivision}
In this section, we describe an integration strategy that subdivides the contour into multiple segments, based on the requirement that the series solutions on each line segment are at most evaluated at a fixed fraction of the distance to the nearest singularity of the differential equations. We call this strategy the predivision strategy, because with this strategy all line segments may be obtained in advance (before doing any expansions.) 

We will work with the set $X_\text{sing}'$ of projections of the singularities on the real line, defined at the end of Section \ref{sec:dynseg}. Next, we define the analogue of Eqns. (\ref{eq:intytildex}) and (\ref{eq:intervx}) by:
\begin{align}
    I(y_{\tilde{x}}) &= [-1/k,1/k]\,, && I(\tilde{x}) = [\tilde{x}^{L}, \tilde{x}^{R}]\,,
\end{align}
such that
\begin{align}
    \tilde{x}^{L} = x(y_{\tilde{x}} = -1/k) && \tilde{x}^{R} = x(y_{\tilde{x}} = 1/k)\,,
\end{align}
and where $k$ is some real number greater than one. 

Next, suppose that we are given boundary conditions $\vec{f}(x_{\text{bc}})$ at $x_{\text{bc}}$, and the point $x_\text{xp}$ on which to center the current line segment. We perform the following three steps, which are very close to those in Section \ref{sec:dynseg}:
\begin{itemize}
    \item[1.] Expand the differential equations in $y_{x_{\text{xp}}}$ and find the corresponding series solutions. Fix boundary conditions at the point $y_{x_{\text{xp}},\text{bc}}$, where $x(y_{x_{\text{xp}}} = y_{x_{\text{xp}},\text{bc}}) = x_{\text{bc}}$.
    \item[2.] If $x_{\text{xp}}^{R} > x_{\text{end}}$, evaluate the series solutions at $y_{x_{\text{xp}},\text{end}}$, where $x(y_{x_{\text{xp}}} = y_{x_{\text{xp}},\text{end}}) = x_{\text{end}}$, and return the result. Otherwise, let $x_{\text{bc}}' = x_{\text{xp}}^{R}$, and evaluate the series solutions at $y_{x_{\text{xp}}}=1/k$, to obtain the next set of boundary conditions $\vec{f}(x_{\text{bc}}')$.
    \item[3.] If $x_{\text{bc}}' \in I(x_i)$ for some $x_i \in X_{\text{sing}}'$, then let $x_{\text{xp}}' = x_i$. Otherwise, let $x_{\text{xp}}'$ be the point such that $x_{\text{xp}}^{\prime, L} = x_{\text{bc}}'$.
\end{itemize}
The third step differs from the one in Section \ref{sec:dynseg} in an important way. Instead of letting $x_{\text{xp}}^{\prime} = x_{\text{bc}}'$, we define $x_{\text{xp}}^{\prime}$ as the point for which $x_{\text{bc}}'$ lies on the left boundary of the interval $I(x_{\text{xp}}')$. This way, we are able to cover more distance with less line segments. In the dynamic strategy we are not able to solve this condition efficiently, as computing the interval $I^{\delta}(x_{\text{xp}}')$ requires expanding the differential equations at $x_{\text{xp}}'$. However, in the current scenario, we may algebraically solve the equation
\begin{align}
    x(y_{x_{\text{xp}}'} = -1/k) = x_{\text{bc}}'\,.
\end{align}
If we use straight line segments we have:
\begin{align}
    x_{\text{xp}}' = x_{b c}^{\prime}+\frac{s}{k}\,
\end{align}
where $s$ is the distance of $x_{\text{xp}}'$ to the nearest singularity, which is given by:
\begin{align}
    s = \left\{\begin{array}{ll}
        \frac{k\left(x_{b c}'-\tilde{x}_{L}\right)}{-1+k} & \text { if } \tilde{x}_{L}<x_{b c}^{\prime}<\frac{\tilde{x}_{L}(1+k)+\tilde{x}_{R}(k-1)}{2 k} \\
        \frac{k\left(-x_{b c}^{\prime}+\tilde{x}_{R}\right)}{1+k} & \text { if } \frac{\tilde{x}_{L}(1+k)+\tilde{x}_{R}(k-1)}{2 k} \leq x_{b c}'<\tilde{x}_{R}
    \end{array}\right.
\end{align}
If we use the Möbius transformed line segments of Eq. (\ref{eq:lineparsmobius}), we have simply:
\begin{align}
    x_{\text{xp}}' = \frac{2 \tilde{x}_{L} \tilde{x}_{R}+x_{b c}^{\prime}\left[(-1+k) \tilde{x}_{L}-(1+k) \tilde{x}_{R}\right]}{-2 x_{b c}^{\prime}+(1+k) \tilde{x}_{L}-(-1+k) \tilde{x}_{R}}\,.
\end{align}
We can take limits of the above equation when $\tilde{x}_L = -\infty$ or when $\tilde{x}_R = +\infty$. We find that the predivision strategy typically needs less line segments than the dynamic integration strategy, in order to obtain results at a given precision. The predivision strategy is enabled in DiffExp by default, with the variable $k$ set to 2, which is controlled by the configuration option \texttt{DivisionOrder}.

\section{The DiffExp package}
\label{sec:DiffExp}
The DiffExp Mathematica package is the main contribution of this paper. The latest version can be downloaded from \href{https://gitlab.com/hiddingm/diffexp}{\color{airforceblue}https://gitlab.com/hiddingm/diffexp}. DiffExp can be loaded into Mathematica using the \texttt{Get[...]}, command, i.e.:
\begin{minted}[linenos=false]{text}
<< "DiffExp.m";
\end{minted}
Note that DiffExp has been designed and tested on Mathematica 12.1. We describe the functions implemented in DiffExp next.

\subsection{Main functions}
\mintinline{text}{LoadConfiguration[l_List]} / \mintinline{text}{UpdateConfiguration[l_List]} / 
\mintinline{text}{UpdateConfiguration[l__Rule]}~\\
First we should parse the configuration options to DiffExp. This is done using the commands \texttt{LoadConfiguration[...]} or \texttt{UpdateConfiguration[..]}. The commands take in a list of rules of configuration options and their values. The function \texttt{LoadConfiguration[...]} sets default values for options which are not included in the argument, while the function  \texttt{UpdateConfiguration[...]} can be used to change individual configuration options. Most options have default values, as described in the table below. The only option that is mandatory is the option \texttt{MatrixDirectory}, which should be a path to a directory containing the partial derivative matrices. For many practical purposes, the option \texttt{DeltaPrescriptions}, which defines the $i\delta$-prescriptions for the analytic continuation, is also mandatory. If it is not specified, it is still possible to transport boundary conditions within a region where no physical threshold singularities are crossed. The full list of options is described next.
\begin{tabularx}{\textwidth}{ | l | X | l |} 
\hline
Option & Description & Default \\\hline
\multicolumn{3}{|c|}{Main configuration options \xrowht[()]{10pt}} \\ \hline
\texttt{DeltaPrescriptions} & A list of polynomials in the kinematic invariants and internal masses, each of which should contain an explicit factor $\pm i\delta$. The zeros of the polynomials should describe singularities such as physical threshold singularities, or branch points of square roots. & \{\} \\\hline
%"EstimateError" & Determines whether to provide error estimates of the results of \texttt{TransportTo[...]}. Possible values are "Precise", "Fast" and False.  & "Precise" \\\hline
\texttt{EpsilonOrder} & An integer specifying the highest order in the dimensional regulator $\epsilon$ in which the integrals should be computed. & 4 \\\hline
\texttt{LineParameter} & The line parameter used for parsing lines to DiffExp. & $x$ \\\hline
\texttt{MatrixDirectory} & The location of a directory on the file system which contains the partial derivative matrices $A_{s}^{(k)}$. The files should be named according to the convention: \texttt{ds\_k.m}, where $s$ is an external scale or a mass variable, and where $k$ is the order in $\epsilon$. A special file \texttt{d\_1.m} may be provided for a canonical polylogarithmic family, which should contain a matrix whose entries are $\mathbb{Q}$-linear combinations of logarithms (the alphabet letters.) &  \\\hline
\texttt{Variables} & The kinematic invariants and masses of the family of basis integrals. If no value is provided, DiffExp will attempt to load all files with the name \texttt{d*\_*.m} at the location specified by the option \texttt{MatrixDirectory}. & \{\}\\\hline
\multicolumn{3}{|c|}{Options related to precision and numerics\xrowht[()]{10pt}}  \\ \hline
\texttt{AccuracyGoal} & The option \texttt{AccuracyGoal} can be used to control the precision of the results. This option is required when the dynamic segmentation strategy is used, and is optional for the predivision strategy. There are a few limitations, discussed below this table.
& - \\\hline
\texttt{ChopPrecision} & Indicates the number off zeros after the decimal point after which terms should be set to 0 in intermediate computations. & 250 \\\hline
\texttt{DivisionOrder} & This option determines the inverse distance to the nearest singularity at which the line segments are evaluated, when the predivision strategy is used. & 3 \\\hline
\texttt{ExpansionOrder} & Specifies the maximum power of the line parameter that should be kept in intermediate series expansions. At intermediate steps the expansions might be multiplied by poles, and the final results may be provided at a lower expansion order. & 50 \\\hline
\texttt{RadiusOfConvergence} & This option has the effect of rescaling the line parameter of each line segment, so that the minimal radius of convergence is given by its value. Higher values may help to combat fastly growing series coefficients. & 1 \\\hline
\texttt{SegmentationStrategy} & This option determines which segmentation strategy is used. The possible values are \texttt{"Dynamic"} and \texttt{"Predivision"}. & \texttt{"Predivision"} \\\hline
\texttt{IntegrationStrategy} & Determines how the differential equations are solved. The value \texttt{"Default"} corresponds to the strategy of Section \ref{sec:gensols}, and is the fastest. The value \texttt{"VOP"} corresponds to using variation of parameters, described in Section \ref{sec:variationofparameters}. This strategy is generally a bit slower for solving coupled integrals, but works along degenerate lines. & \texttt{"Default"} \\\hline
\texttt{UseMobius} & This option determines whether the line segments are obtained by linear transformations or by Möbius transformations. & False \\\hline
\texttt{UsePade} & Determines whether Padé approximants are used while transporting boundary conditions. & False \\\hline
\texttt{WorkingPrecision} & The number of digits kept for any intermediate computation. & 500 \\\hline
\multicolumn{3}{|c|}{Other options \xrowht[()]{10pt}}  \\\hline 
\texttt{LogFile} & Location of a log file on which to write all output of the current session. & - \\\hline
\texttt{Verbosity} & Determines the level of printed output. The default level is 1 and the maximum level is 3. When running inside a Mathematica notebook, lower verbosity levels are generally recommended. For shell-scripts higher verbosity levels might be preferred. & 1\\\hline
\end{tabularx}
We provide additional comments about some of the configuration options next.
{
\paragraph{\texttt{AccuracyGoal}:}
\parindent = 0.7cm \hangindent=0.7cm 
When \texttt{AccuracyGoal} is specified, DiffExp will aim to transport the boundary conditions at an absolute precision of $10^{-\delta}$, where $\delta$ is the value of \texttt{AccuracyGoal}. The option \texttt{AccuracyGoal} works by bounding the error of the expansions of the differential equations. For integrals that are not coupled, or coupled at low orders, it is typically the case that the solutions to the differential equations have the same error. For highly coupled sectors we did not always find this to be the case, and in the presence of such sectors setting the option \texttt{AccuracyGoal} might not have the desired effect. In this case, one may still increase or lower the value of \texttt{AccuracyGoal} to control the precision globally. By default, \texttt{AccuracyGoal} is turned off, and the precision can be controlled using the options \texttt{ExpansionOrder} and \texttt{DivisionOrder}. 

\parindent = 0.7cm \hangindent=0.7cm 
If the option \texttt{SegmentationStrategy} is set to \texttt{"Dynamic"}, the option \texttt{AccuracyGoal} determines how far the solutions are evaluated away from the origin. If the option \texttt{SegmentationStrategy} is set to \texttt{"Predivision"}, DiffExp will dynamically increase or decrease the expansion order of each line segment, until the expansions of the differential equations are within the desired precision. This means that the differential equations are expanded multiple times, until the desired precision is reached. If the expansion of the differential equations bottlenecks the computation, then using the option \texttt{AccuracyGoal} with the predivision segmentation strategy is not recommended.

\parindent = 0.7cm \hangindent=0.7cm Lastly, note that \texttt{AccuracyGoal} does not take into account Padé approximants in determining the error. When \texttt{AccuracyGoal} is specified, and Padé approximants are enabled, the precision of the results are typically far higher than the given \texttt{AccuracyGoal}. In this case, setting a value for \texttt{AccuracyGoal} might still be useful for globally increasing or decreasing the precision.

%\paragraph{"EstimateError"} There are two ways that DiffExp can estimate the error of the result. If the value of "EstimateError" is set to "Precise", DiffExp transports the boundary conditions twice at different expansion orders. This typically gives a good estimate of the error, but slows down the computation of the integrals by at most a factor of two. Typically the additional time needed for computing the errors is much less, as costly parts in the computation of the coupled sectors are recycled, and so are the expansion of the partial derivative matrices. If the value of "EstimateError" is set to "Fast", the reported error is simply read of from the highest expansion coefficients of the solutions. The estimated error typically gets closer to the true error, the more line segments are computed.
}

{
\texttt{ChopPrecision}, \texttt{RadiusOfConvergence}, and \texttt{WorkingPrecision}{\bf : }
\parindent = 0.7cm \hangindent=0.7cm
We provide a number of comments about these three options, which impact the numerical precision and stability of the calculations. Firstly, the option \texttt{WorkingPrecision} determines the number of digits at which inexact numbers are kept. The value of \texttt{WorkingPrecision} should typically be put significantly higher than the precision that is desired for the final results. This is because at intermediate stages there might for example be cancellations between large numbers. The value of \texttt{ChopPrecision} determines the number off zeros after the decimal point after which numbers are discarded. For families of integrals where the number of coupled integrals is low, such as a polylogarithmic family in a canonical basis, the value of \texttt{ChopPrecision} can be set a bit lower than the value of \texttt{WorkingPrecision}. For integral families where there are sectors with many coupled integrals, the value of \texttt{ChopPrecision} may need to be set significantly lower than the value of \texttt{WorkingPrecision}. One reason for this is that solving the coupled sectors involves calls to a number of Mathematica's linear algebra routines, for which the options \texttt{Tolerance} and \texttt{ZeroTest} are controlled by the value of \texttt{ChopPrecision}. If very small nonzero numbers remain present in the matrices, the linear algebra routines may run into numerical instabilities. Note that by default the values of \texttt{ChopPrecision} and \texttt{WorkingPrecision} are set fairly high, so that most problems simply run out of the box. One can look at the example notebooks to see some other typical configuration values.

\parindent = 0.7cm \hangindent=0.7cm
Lastly, another option that may affect the numerical stability is \texttt{RadiusOfConvergence}. It has the effect of rescaling all series coefficients in the manner $c_k x^k \rightarrow c_k (x/10)^k$. This may be useful when the expansions blow up at intermediate stages. By default the value of \texttt{RadiusOfConvergence} is set to one. In the three-loop unequal-mass banana graph family, we found it necessary to set this option higher than one, in order to obtain stable numerical behaviour. For all other examples we didn't need to use this option. The three-loop unequal-mass banana graph family is currently somewhat of an edge case for DiffExp, since it involves a sector of eleven coupled integrals. Also note that setting the value of \texttt{RadiusOfConvergence} too high may result in the expansion coefficients becoming too small, and being incorrectly discarded, at intermediate stages of the calculation.

}

{
\paragraph{\texttt{DeltaPrescriptions}:} 
\parindent = 0.7cm \hangindent=0.7cm
The option \texttt{DeltaPrescriptions} should be given a list of polynomials with associated $i\delta$-prescriptions, such that the zero sets of the polynomials correspond to physical threshold singularities, or the arguments of square roots in the basis choice. In order to find results at any given point in phase-space, the list should contain all the physical threshold singularities of the basis integrals. In practice, one only has to provide the physical threshold singularities that need to be crossed. For example, if the boundary conditions are provided in the Euclidean region, one would provide the necessary $i\delta$-prescriptions to analytically continue the results to the physical region of interest. By default, DiffExp will recognize which square roots appear in the differential equations, and assign them the $+i\delta$ prescription (i.e. the principal branch), unless otherwise specified.

\parindent = 0.7cm \hangindent=0.7cm
There are two equivalent ways that the $i\delta$-prescriptions may be passed to DiffExp. The first way involves adding explicit terms of the form $\pm i\delta$ to the polynomials, while the second method involved adding the signs of the $i\delta$-prescription as a separate argument. For example, we could provide either of the following:
\begin{minted}[linenos=false]{text}
    DeltaPrescriptions -> {-s+4-Iδ, t-4+Iδ},
    DeltaPrescriptions -> {{-s+4,-1}, {t-4,1}},
\end{minted}

\parindent = 0.7cm \hangindent=0.7cm
to define the prescriptions for threshold singularities at $t = 4$ and $s=4$.

\parindent = 0.7cm \hangindent=0.7cm
Lastly, we mention a potential pitfall regarding the analytic continuation. For each line segment, DiffExp checks whether there are multivalued functions in the expansions. If multivalued functions are present, but DiffExp is not centered at one of the singular regions provided by \texttt{DeltaPrescriptions}, the computation will be aborted and DiffExp will ask the user to provide the relevant $i\delta$-prescription. One situation where this check fails, is if two singular regions intersect at the origin of the line segment but the analytic continuation prescription is only given for one of them. In this case, DiffExp will assume the $i\delta$-prescription of the singular region that was provided to the option \texttt{DeltaPrescriptions}, which might not be the correct choice for the other one.

}

{
\paragraph{\texttt{MatrixDirectory}:} 
\parindent = 0.7cm \hangindent=0.7cm
The partial derivative matrices that are provided to DiffExp may only contain combinations of rational functions and square roots with irreducible polynomial arguments. Other functions such as elliptic integrals, which show up for canonical bases of elliptic families of Feynman integrals, are not supported. However, for such families one may provide a precanonical basis instead. If a file \texttt{ds\_k.m} is absent, for some epsilon order $k$, it is assumed that the corresponding matrix has all entries equal to zero. To speed up the expansions of polylogarithmic sectors, a special matrix \texttt{d\_1.m} may be provided, whose entries should be linear combinations of logarithms. Note that if both \texttt{d\_1.m}, and files of the form \texttt{ds\_1.m}, are present in the folder, their contributions will be summed together.

}

{
\paragraph{{ \tt UseMobius}:} 
\parindent = 0.7cm \hangindent=0.7cm
Enabling Möbius transformations reduces the number of line segments needed to transport boundary conditions at a given precision. However, if we work on Möbius transformed line segments, the time needed for expanding the differential equations might increase considerably. When the differential equations are large, their expansion might be the main computational bottleneck, and in such cases enabling Möbius transformations can be detrimental to performance. For this reason, Möbius transformations are turned off by default.

}

{
\paragraph{\texttt{UsePade}:}
\parindent = 0.7cm \hangindent=0.7cm
Enabling Padé approximants typically increases the precision of the solutions considerably. However, the use of Padé approximants can sometimes lead to numerical instabilities. This is typically the case when the options \texttt{ChopPrecision} and \texttt{WorkingPrecision} are set to values that are too low. Furthermore, finding the Padé approximants is somewhat costly too, and adds computation time to the algorithm. Typically, we still find that the use of Padé approximants decreases the computation time needed to obtain results at a given precision, but to be safe, they are currently turned off by default.

}

\paragraph{\mintinline{text}{CurrentConfiguration[]}}~\\
This function return a list with the current configuration options.
\paragraph{\mintinline{text}{PrepareBoundaryConditions[bcs_List, line_List]}}~\\
This function converts a set of boundary conditions into a form that is useable by the routines \texttt{IntegrateSystem[...]} and \texttt{TransportTo[...]}. The first argument should contain the boundary conditions, while the second argument should contain a point or line specifying an asymptotic limit in phase-space, in which the boundary conditions are given. DiffExp recognizes whether the argument is a line or a point, by checking whether it depends on the line parameter.
 
The first argument should be a list of $n$ elements, which contain the boundary conditions of the integrals. The boundary conditions of an individual integral can be given in one of the following three forms:
\begin{itemize}
  \item[1.] A closed-form expression in $\epsilon$.
  \item[2.] A list of coefficients for each order in $\epsilon$, where the first list element corresponds to order $\epsilon^0$.
  \item[3.] The string \texttt{"?"}, which instructs DiffExp to ignore boundary conditions for the integral. This option is useful for when dealing with coupled integrals in an asymptotic limit, where the boundary conditions for a subset of the integrals may fix the remaining ones.
\end{itemize}

If the second argument is a line, DiffExp assumes that the boundary conditions given in the first argument are valid at leading order in the limit where the line parameter $x$ approaches the origin from the positive direction. More specifically, if the leading order is proportional to $x^k$, then DiffExp will assume the boundary conditions are valid up to $\mathcal{O}(x^{k+1/2})$. To override this behaviour, one may provide the boundary conditions as a list of terms, order-by-order in $\epsilon$, where each term itself is a series expansion in $x$ (given by Mathematica's \texttt{SeriesData} object, i.e. the output of the \texttt{Series[...]} function.) DiffExp will then assume the results to be valid up to the order to which the series is provided.

If the boundary conditions contain multivalued functions, which is typical for asymptotic limits, they should be provided in such a way that the positive direction of the line across which the limit is taken points along the standard (Mathematica branch) of the multivalued function. For example, suppose that the boundary conditions contain a term of the form $\log(-s)$, and that the Feynman prescription dictates that $s$ should carry a positive imaginary part. This situation will lead to incorrect results along the line $s=x$, since DiffExp will convert the logarithm into the form
\begin{align}
    \log(-s(x)) = i \pi + \log(x)\,.
\end{align}
The correct way to pass the boundary term to DiffExp is therefore to change $\log(-s)$ to $-i \pi + \log(s)$ before calling \texttt{PrepareBoundaryConditions[...]}. Similar considerations apply when passing closed-form expressions like $(-s)^\epsilon$.

Note that the output of \texttt{PrepareBoundaryConditions[...]} includes the point or line that was given in the second argument. That way, when feeding the result to \texttt{IntegrateSystem[...]} or \texttt{TransportTo[...]}, DiffExp knows where to fix the boundary conditions.
\paragraph{\mintinline{text}{IntegrateSystem[bcs_List, line_List]}} ~\\The function \texttt{IntegrateSystem[...]} implements the integration of the differential equations along a single line segment. It is possible to omit the first argument, and \texttt{IntegrateSystem[...]} will then return the general solution to the differential equations at the given point. The free parameters will be labelled using the convention $c_{i,j,k}$, where $i$ corresponds to the order in $\epsilon$, $j$ to the coupled block of integrals, and $k$ labels the parameters.

When boundary conditions are provided, the first argument should be the output of the function \texttt{PrepareBoundaryConditions[...]}, or the output of the function \texttt{TransportTo[...]}. If the boundary conditions are given at a point, the point should lie on the line given as the second argument. If the boundary conditions are given as an asymptotic limit, the line along which the boundary conditions are given should be parallel, oriented in the same direction, and centered at the line passed to \texttt{IntegrateSystem[...]}. If the two lines satisfy these conditions, but were parametrized differently, DiffExp will automatically perform the change of parametrization in the boundary terms.

The output of \texttt{IntegrateSystem[...]} is an $(n\times m)$-matrix, where $n$ is the number of basis integrals and where $m$ is equal to the value of the option \texttt{EpsilonOrder} plus one. The first column gives the $\epsilon^0$-coefficients of the integrals. If the expansions were centered at a branch point, the result may contain the Heaviside step functions $\theta(x)$ and $\theta(-x)$, which are labelled as $\theta_p$ and $\theta_m$ respectively.

\mintinline[breaklines]{text}{TransportTo[bcs_List, line_List, to_:1, save_:False]}~\\
The function \texttt{TransportTo[...]} is the most important function in DiffExp, as it performs the transportation of boundary conditions to arbitrary (real-valued) points in the phase-space of kinematic invariants and internal masses. The conditions on the arguments \texttt{bcs\_} and \texttt{line\_} are the same as for the function \texttt{IntegrateSystem[...]}, in the case that \texttt{line\_} depends on the line parameter $x$. The results will then be transported to the endpoint \texttt{line /. x $\rightarrow$ to}. If the argument \texttt{line\_} is a point instead, DiffExp will consider the line \texttt{x*line + (1-x)*start}, where \texttt{start} is the point at which the boundary conditions were prepared using \texttt{PrepareBoundaryConditions[...]}.

The argument \texttt{save\_} determines whether the expansions along individual line segments should be saved and returned in the output. If it is set to true, the output of \texttt{TransportTo[...]} may be passed to the function \texttt{ToPiecewise[...]}, which combines the results of all line segments together into a single function, which is suitable for numerical evaluation, or for plotting purposes. If the argument \texttt{save\_} is set to false, the output of \texttt{TransportTo[...]} is a list consisting of the form \{point, results, errors\}. The first list element is the point in phase-space at which the results were evaluated. The second and third element of the list are both $(n\times m)$-matrices, where $n$ is the number of basis integrals and where $m$ is equal to the value of the option \texttt{EpsilonOrder} plus one. %The third element of the list is only returned when the option "EstimateError" is set to true. 
If the argument \texttt{save\_} is set to true, the output of \texttt{TransportTo[...]} has instead the form \{\{point, results, errors\}, segmentdata\}, where segment data is a list which encodes the expansions obtained along individual line segments.

The error estimates are provided as a convenience to the user, but should probably not be relied upon for sensitive results. In that case, a better way to estimate the error, is to evaluate a point along two different contours, and to take the difference between the results. The error estimates are obtained in the following way. At each matching point between neighbouring line segments, and at the final evaluation point, we also evaluate the series solutions at an order that is reduced by a certain number $q > 0$. We then compute the difference between the evaluation of the lower order solutions and the original solutions, and take the absolute value. The number $q$ is currently determined by a simple heuristic. In particular, we found that it was useful to let $q$ be proportional to the maximum order at which integrals are coupled in the integral family, in order to get reliable estimates for highly coupled families. The error accumulated along each line segment is added to the total error estimate. Note that if the option \texttt{UsePade} is set to true, the evaluation of the lower order series solutions is also done using Padé approximants.

\paragraph{\mintinline{text}{ToPiecewise[segmentdata_List, pade_:False]}}
The function \texttt{ToPiecewise[...]} takes as input the output of \texttt{TransportTo[...]}, given that the latter has been run with the argument \texttt{save\_} equal to true. The output of \texttt{ToPiecewise[...]} is an $(n\times m)$-matrix, where $n$ is the number of basis integrals and where $m$ is equal to the value of the option \texttt{EpsilonOrder} plus one. Each entry is a \texttt{Piecewise} mathematica object, which is a function of the line parameter of the line that was given to \texttt{TransportTo[...]}. The output of \texttt{ToPiecewise[...]} may be used for numerical evaluation of the results at arbitrary points along the line, or for plotting purposes.

The argument \texttt{pade\_} determines whether the \texttt{Piecewise} objects are composed out of the Padé approximants of the solutions along the line segments, or out of the series solutions. If \texttt{TransportTo[...]} was called with the configuration option \texttt{UsePade} to false, there should not be a significant difference in precision by enabling Padé approximants here. Note that computing the Padé approximants might take some time, and if one is just interested in plotting results then it is usually not necessary to compute the Padé approximants. However, if the aim is to use the output of \texttt{ToPiecewise} for numerical evaluation, it is advised to set \texttt{pade\_} to true.

\section{Examples}
\label{sec:examples}
In the following section we consider two examples in detail, the equal-mass three-loop banana family, and its unequal mass generalization. The results in this section can be obtained by running the notebook \texttt{Banana.nb} in the \texttt{Examples} folder shipped with DiffExp. We discuss a few other examples at the end of this section.

\subsection{Equal-mass three-loop banana family}
\begin{figure}[h]
    \centering
    \includegraphics[width=0.3\textwidth]{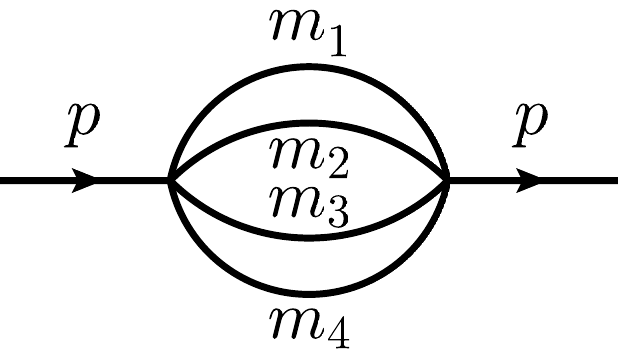}
    \caption{The three-loop unequal mass banana diagram.}
    \label{fig:unequalmassbanana}
\end{figure}
\begin{figure}[H]
\centering
\includegraphics[width=12cm]{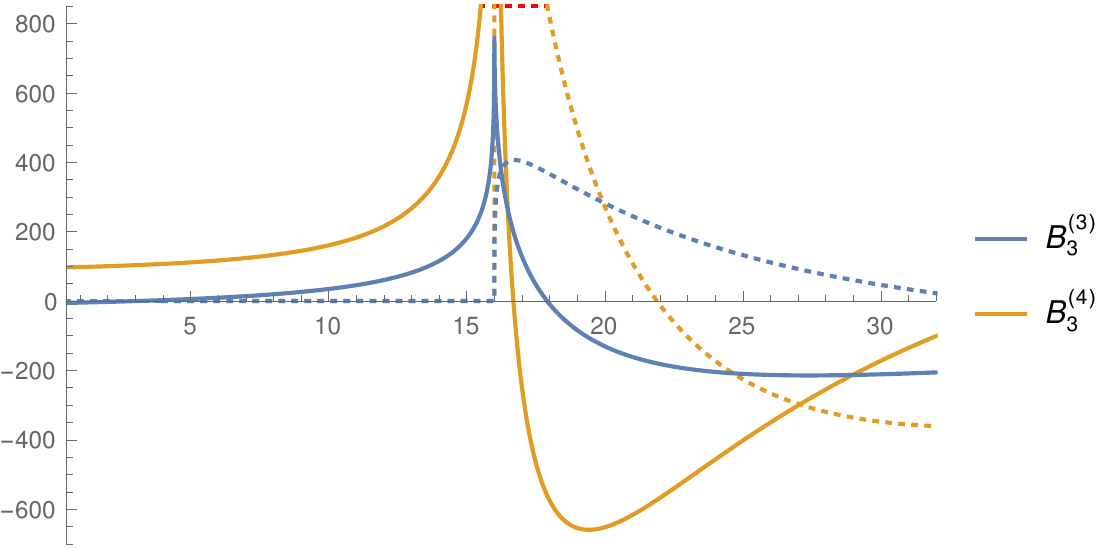}
\caption{Plot of the master integral $B_3$ in the region $p^2/m^2 = 0 \ldots 32$. The solid lines are the real parts of the integrals, and the dotted lines the imaginary parts.}
\label{fig:figequalmassplot}
\end{figure}

The three-loop unequal-mass banana diagram is depicted in Fig. \ref{fig:unequalmassbanana}. We will first consider the equal-mass case, in which we let $m_i^2 = m^2$ for $i = 1, \ldots, 4$. We will normalize out the overall mass dimension, and parametrize the kinematics by the ratio $t=p_{1}^{2} / m^{2}$. Furthermore, we will work in the dimension $d = 2-2\epsilon$. We define the equal-mass banana integral family by:
\begin{align}
    \label{eq:bananafidefinition}
     I^{\text{banana}}_{a_1a_2a_3a_4} &= \left(\frac{ e^{\gamma_E \epsilon}}{i \pi^{d/2}} \right)^3 (m^2)^{a-\frac{3}{2}(2-2\epsilon)}\left(\prod_{i=1}^4 \int d^dk_i\right) D_1^{-a_1} D_2^{-a_2} D_3^{-a_3} D_4^{-a_4}\,.
\end{align}
where the propagators are:
\begin{align}
    D_1 &= -k_1^2 + m^2\,, & D_2 &= -k_2^2 + m^2\,, \nonumber\\
    D_3 &= -k_3^2 + m^2\,, & D_4 &= -(k_1+k_2+k_3+p_1)^2 + m^2\,.
\end{align}
For brevity, we have not included numerator terms in the definition of the integral family, because we can obtain a basis of master integrals without numerators for this integral family. We choose the basis of master integrals to be:
\begin{align}
    \vec{B}^{\text{banana}} = (\epsilon I^{\text{banana}}_{2211},\, \epsilon (1+3\epsilon) I^{\text{banana}}_{2111},\, \epsilon (1+3 \epsilon)(1+4 \epsilon) I^{\text{banana}}_{1111},\, \epsilon^3 I^{\text{banana}}_{1110})\,,
\end{align}
for which the differential equations are in precanonical form. They are given by:
\begin{equation}
\partial_t\vec{B}^{\text{banana}} = 
\left(
\begin{array}{cccc}
 -\frac{64-2 t+t^{2}+(8+t)^{2} \epsilon}{t(t-16)(t-4)} & \frac{2 (t+20) (2 \epsilon +1)}{t(t-16) (t-4)} & -\frac{6 (2 \epsilon +1)}{t(t-16) (t-4)} & -\frac{2 \epsilon }{t(t-16)} \\
 \frac{3 t (3 \epsilon +1)}{t(t-4)} & -\frac{2(t+8) \epsilon+t+4}{t(t-4)} & \frac{3 \epsilon +1}{t(t-4)} & 0 \\
 0 & \frac{4 (4 \epsilon +1)}{t} & \frac{-3 \epsilon -1}{t} & 0 \\
 0 & 0 & 0 & 0 \\
\end{array}
\right)\vec{B}^{\text{banana}}
\end{equation}
The IBP reductions required for setting up the differential equations were obtained using \texttt{Kira} \cite{Maierhoefer:2017hyi, Maierhofer:2018gpa, Klappert:2020nbg}. We seek to compute boundary conditions for the system of differential equations. Since the first three master integrals are coupled, it turns out we only have to provide boundary conditions for the master integrals $I_{1111}^\text{banana}$ and $I_{1110}^\text{banana}$, the latter of which is trivial and given by:
\begin{align}
    I_{1110}^{\text{banana}} = e^{3 \gamma \epsilon} \Gamma(\epsilon)^{3}\,.
\end{align}
We will compute boundary conditions for $I_{1111}^\text{banana}$ in the limit $t = -1/x$, with $x\downarrow 0$. We will occasionally refer to this as the infinite momentum limit. The Feynman parametrization of $I_{1111}^{\text{banana}}$ is given by:
\begin{align}
    \begin{aligned}
I_{1111}^{\text{banana}} =  &\,\, i e^{3 \gamma \epsilon}  \Gamma(3 \epsilon+1) \left(m^{2}\right)^{-3 \epsilon-1} x^{3 \epsilon+1} \int_{\Delta^3}\left[d^3\vec{\alpha}\right]\left(\alpha_{1} \alpha_{2} \alpha_{3}+\alpha_{1} \alpha_{4} \alpha_{3}+\alpha_{2} \alpha_{4} \alpha_{3}+\right.\\
&\left.\alpha_{1} \alpha_{2} \alpha_{4}\right)^{4 \epsilon} \left(\alpha_{2} \alpha_{3} \alpha_{1}^{2} x+\alpha_{2} \alpha_{4} \alpha_{1}^{2} x+\alpha_{3} \alpha_{4} \alpha_{1}^{2} x+\alpha_{2} \alpha_{3}^{2} \alpha_{1} x+\alpha_{2} \alpha_{4}^{2} \alpha_{1} x+\alpha_{3} \alpha_{4}^{2} \alpha_{1} x+\right.\\
&\left.\alpha_{2}^{2} \alpha_{3} \alpha_{1} x+\alpha_{2}^{2} \alpha_{4} \alpha_{1} x+\alpha_{3}^{2} \alpha_{4} \alpha_{1} x+4 \alpha_{2} \alpha_{3} \alpha_{4} \alpha_{1} x+\alpha_{2} \alpha_{3} \alpha_{4}^{2} x+\alpha_{2} \alpha_{3}^{2} \alpha_{4} x+\alpha_{2}^{2} \alpha_{3} \alpha_{4} x+\right.\\
&\left.+\alpha_{2} \alpha_{3} \alpha_{4} \alpha_{1}\right)^{-3 \epsilon-1}\,.
\end{aligned}
\end{align}
From \texttt{asy}, we obtain fifteen regions as $x\downarrow 0$:
\begin{equation}
\begin{array}{lll}
R_{1}=\{0,-1,-1,-1\}\,, & R_{2}=\{0,-1,-1,0\}\,, & R_{3}=\{0,0,0,0\}\,, \\
R_{4}=\{0,0,0,-1\}\,, & R_{5}=\{0,1,1,0\}\,, & R_{6}=\{0,0,1,0\}\,, \\
R_{7}=\{0,-1,0,-1\}\,, & R_{8}=\{0,-1,0,0\}\,, & R_{9}=\{0,0,0,1\}\,, \\
R_{10}=\{0,1,1,1\}\,, & R_{11}=\{0,0,1,1\}\,, & R_{12}=\{0,1,0,0\}\,, \\
R_{13}=\{0,0,-1,-1\}\,, & R_{14}=\{0,1,0,1\}\,, & R_{15}=\{0,0,-1,0\}\,.
\end{array}
\end{equation}
At leading order in $x$ and in each region $R_i$, the resulting parametric representation for $I_{1111}^{\text{banana}}$ may be integrated directly. The contributions of all regions are given by:
\begin{align}
\begin{array}{lll}
 \text{I}_{1111}^{R_1}\sim x e^{3 \gamma  \epsilon } \Gamma (\epsilon )^3\,, & \text{I}_{1111}^{R_2}\sim \frac{e^{3 \gamma  \epsilon } \epsilon  x^{\epsilon +1} \Gamma (-\epsilon )^2 \Gamma (\epsilon )^3}{\Gamma (-2 \epsilon )}\,, & \text{I}_{1111}^{R_3}\sim \frac{3 e^{3 \gamma  \epsilon } \epsilon  x^{3 \epsilon +1} \Gamma (-\epsilon )^4 \Gamma (3 \epsilon )}{\Gamma (-4 \epsilon )}\,, \\
 \text{I}_{1111}^{R_4}\sim \frac{2 e^{3 \gamma  \epsilon } \epsilon  x^{2 \epsilon +1} \Gamma (-\epsilon )^3 \Gamma (\epsilon ) \Gamma (2 \epsilon )}{\Gamma (-3 \epsilon )}\,, & \text{I}_{1111}^{R_5}\sim \frac{e^{3 \gamma  \epsilon } \epsilon  x^{\epsilon +1} \Gamma (-\epsilon )^2 \Gamma (\epsilon )^3}{\Gamma (-2 \epsilon )}\,, & \text{I}_{1111}^{R_6}\sim x e^{3 \gamma  \epsilon } \Gamma (\epsilon )^3\,, \\
 \text{I}_{1111}^{R_7}\sim \frac{e^{3 \gamma  \epsilon } \epsilon  x^{\epsilon +1} \Gamma (-\epsilon )^2 \Gamma (\epsilon )^3}{\Gamma (-2 \epsilon )}\,, & \text{I}_{1111}^{R_8}\sim \frac{2 e^{3 \gamma  \epsilon } \epsilon  x^{2 \epsilon +1} \Gamma (-\epsilon )^3 \Gamma (\epsilon ) \Gamma (2 \epsilon )}{\Gamma (-3 \epsilon )}\,, & \text{I}_{1111}^{R_9}\sim x e^{3 \gamma  \epsilon } \Gamma (\epsilon )^3\,, \\
 \text{I}_{1111}^{R_{10}}\sim \frac{2 e^{3 \gamma  \epsilon } \epsilon  x^{2 \epsilon +1} \Gamma (-\epsilon )^3 \Gamma (\epsilon ) \Gamma (2 \epsilon )}{\Gamma (-3 \epsilon )}\,, & \text{I}_{1111}^{R_{11}}\sim \frac{e^{3 \gamma  \epsilon } \epsilon  x^{\epsilon +1} \Gamma (-\epsilon )^2 \Gamma (\epsilon )^3}{\Gamma (-2 \epsilon )}\,, & \text{I}_{1111}^{R_{12}}\sim x e^{3 \gamma  \epsilon } \Gamma (\epsilon )^3\,, \\
 \text{I}_{1111}^{R_{13}}\sim \frac{e^{3 \gamma  \epsilon } \epsilon  x^{\epsilon +1} \Gamma (-\epsilon )^2 \Gamma (\epsilon )^3}{\Gamma (-2 \epsilon )}\,, & \text{I}_{1111}^{R_{14}}\sim \frac{e^{3 \gamma  \epsilon } \epsilon  x^{\epsilon +1} \Gamma (-\epsilon )^2 \Gamma (\epsilon )^3}{\Gamma (-2 \epsilon )}\,, & \text{I}_{1111}^{R_{15}}\sim \frac{2 e^{3 \gamma  \epsilon } \epsilon  x^{2 \epsilon +1} \Gamma (-\epsilon )^3 \Gamma (\epsilon ) \Gamma (2 \epsilon )}{\Gamma (-3 \epsilon )}\,. \\
\end{array}
\end{align}
Summing over all the regions, we obtain the final result:
\begin{align}
    I_{1111}^{\text{banana}} &\overset{x\downarrow 0}{\sim} \frac{6 e^{3 \gamma \epsilon} \epsilon x^{\epsilon+1} \Gamma(-\epsilon)^{2} \Gamma(\epsilon)^{3}}{\Gamma(-2 \epsilon)}+\frac{8 e^{3 \gamma \epsilon} \epsilon x^{2 \epsilon+1} \Gamma(-\epsilon)^{3} \Gamma(\epsilon) \Gamma(2 \epsilon)}{\Gamma(-3 \epsilon)}+\frac{3 e^{3 \gamma \epsilon} \epsilon x^{3 \epsilon+1} \Gamma(-\epsilon)^{4} \Gamma(3 \epsilon)}{\Gamma(-4 \epsilon)}\nonumber\\&\quad+4 x e^{3 \gamma \epsilon} \Gamma(\epsilon)^{3} + \mathcal{O}(x^2)\,.
\end{align}
Next, we use DiffExp to plot the banana graph in the region $t = 0,\ldots,32$. First we consider the line $t = -1/x$, and transport the boundary conditions from $x = 0$ to $1$. Thereafter, we transport the result along the line $t=x$, from $-1$ to $32$. The relevant commands are:
\begin{minted}[linenos=false]{text}
Γ = Gamma;
BananaBoundaryConditions = {
    "?", "?",
    ε(1+3ε)(1+4ε)(-((4E^(3εEulerGamma)Γ[ε]^3)/t)+(
        6E^(3εEulerGamma)ε(-(1/t))^(1+ε)Γ[-ε]^2Γ[ε]^3)/
        Γ[-2ε]+(8E^(3εEulerGamma)ε(-(1/t))^(1+2ε)
        Γ[-ε]^3Γ[ε]Γ[2ε])/Γ[-3ε]+
        (3E^(3εEulerGamma)ε(-(1/t))^(1+3ε)Γ[-ε]^4Γ[3ε])/Γ[-4ε]),
    E^(3εEulerGamma)ε^3Γ[ε]^3
}// PrepareBoundaryConditions[#, <|t -> -1/x|>] &;
  
Results1 = TransportTo[BananaBoundaryConditions, <|t -> -1|>];
Results2 = TransportTo[Results1, <|t -> x|>, 32, True];

ResultsFunction = ToPiecewise[Results2];

ReImPlot[{ResultsFunction[[3, 4]][x], ResultsFunction[[3, 5]][x]}, 
    {x, 1/2, 32}, MaxRecursion -> 15, WorkingPrecision -> 100]
\end{minted}
We performed some additional processing of the plot, which gives the result in Fig. \ref{fig:figequalmassplot}. It took about 1 minute to reach the point $p^2/m^2 = 32$ from the limit $p^2/m^2 = -\infty$, with an estimated error of $10^{-25}$, with the option \texttt{DivisionOrder} set to 3, and the option \texttt{ExpansionOrder} set to 50.

\subsection{Unequal-mass three-loop banana family}
\label{sec:fourmassunequalbanana}
Next, we will consider the unequal-mass banana graph family. This time, we will not normalize the integrals by the power of an internal mass. The unequal-mass banana integral family is then defined by:
\begin{align}
    \label{eq:bananauneqdefinition}
    I^{\text{banana}}_{a_1a_2a_3a_4} = \left(\frac{ e^{\gamma_E \epsilon}}{i \pi^{d/2}} \right)^3 \left(\prod_{i=1}^4 \int d^dk_i\right) D_1^{-a_1} D_2^{-a_2} D_3^{-a_3} D_4^{-a_4}\,,
\end{align}
where:
\begin{align}
    D_1 &= -k_1^2 + m_1^2\,, & D_2 &= -k_2^2 + m_2^2\,,\nonumber\\ 
    D_3 &= -k_3^2 + m_3^2\,, & D_4 &= -(k_1+k_2+k_3+p_1)^2 + m_4^2\,.
\end{align}
We choose the following basis of precanonical master integrals:
\begin{align}
    \vec{B}^\text{banana} = \left\{\begin{array}{l}
        \epsilon I_{1122}^\text{banana},\, \epsilon I_{1212}^\text{banana},\, \epsilon I_{1221}^\text{banana},\, \epsilon I_{2112}^\text{banana},\, 
        \epsilon I_{2121}^\text{banana},\, \epsilon I_{2211}^\text{banana},\, \\ \epsilon(1+3\epsilon)I_{1112}^\text{banana},\, \epsilon(1+3\epsilon)I_{1121}^\text{banana},\,
        \epsilon(1+3\epsilon)I_{1211}^\text{banana},\, \\ \epsilon(1+3\epsilon)I_{2111}^\text{banana},\, \epsilon(1+3\epsilon)(1+4\epsilon)I_{1111}^\text{banana},\, \\
        \epsilon^3 I_{0111}^\text{banana},\,
        \epsilon^3 I_{1011}^\text{banana},\,
        \epsilon^3 I_{1101}^\text{banana},\,
        \epsilon^3 I_{1110}^\text{banana}
    \end{array}\right\}\,.
\end{align}
We will label the basis integrals from left to right, and top to bottom, by $B_1$, \ldots, $B_{15}$, and we denote their $\epsilon$-orders by a superscript. The corresponding differential equations are 8 megabytes in size, and too large to present here. The required IBP reductions were obtained using \texttt{Kira}.

The unequal-mass family is significantly more difficult to compute than the equal-mass family, due to the fact that there are eleven coupled integrals in the top sector. Furthermore, we found that at intermediate steps of the calculation the series coefficients are growing very fast with the order of the line parameter. We compensated for this by setting the options \texttt{ChopPrecision} and \texttt{WorkingPrecision} very high, and setting the option \texttt{RadiusOfConvergence} to 10. This has the effect of rescaling all series coefficients in the manner $c_k x^k \rightarrow c_k (x/10)^k$. 

\begin{figure}[h!]
	\begin{tabular}{ p{7cm}  p{7cm}}
        \imgPlot{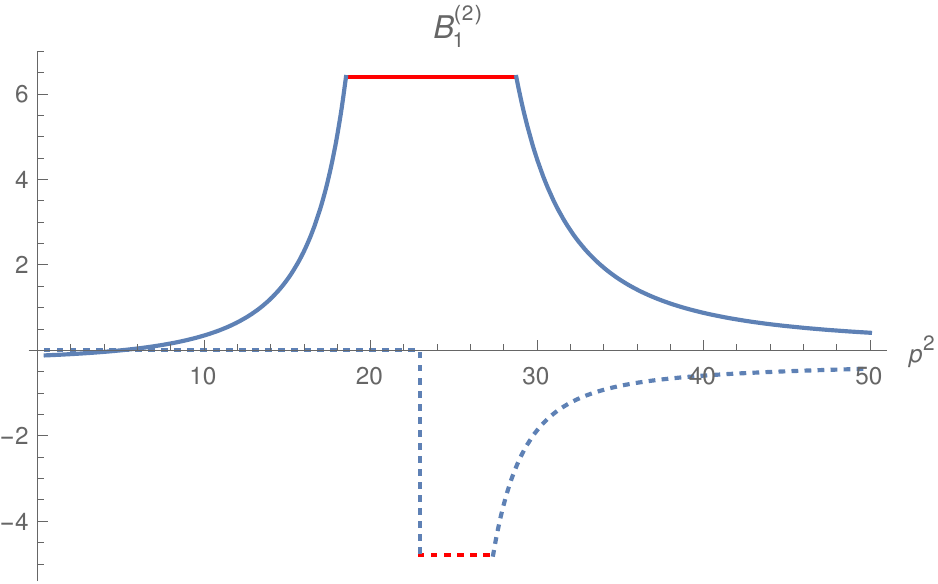} & \imgPlot{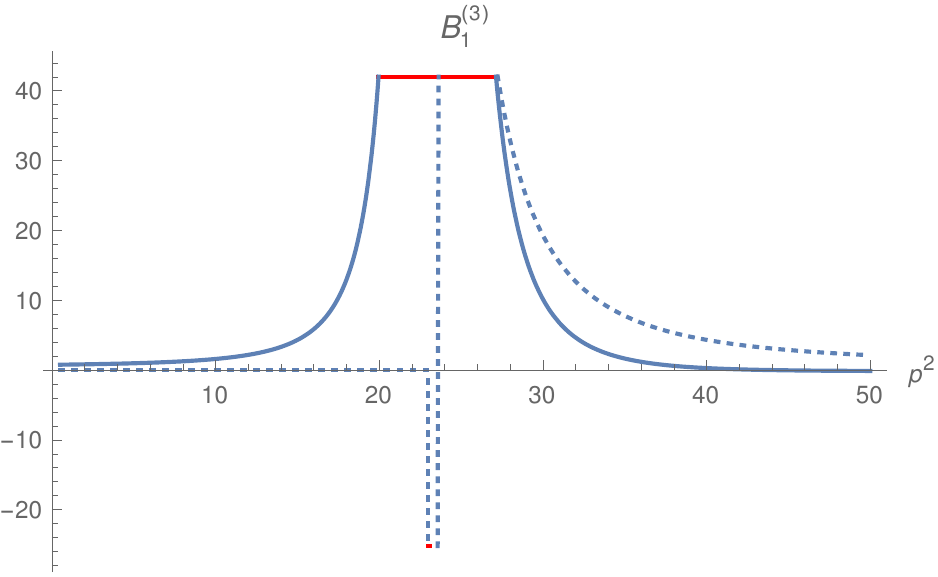} \\
        \imgPlot{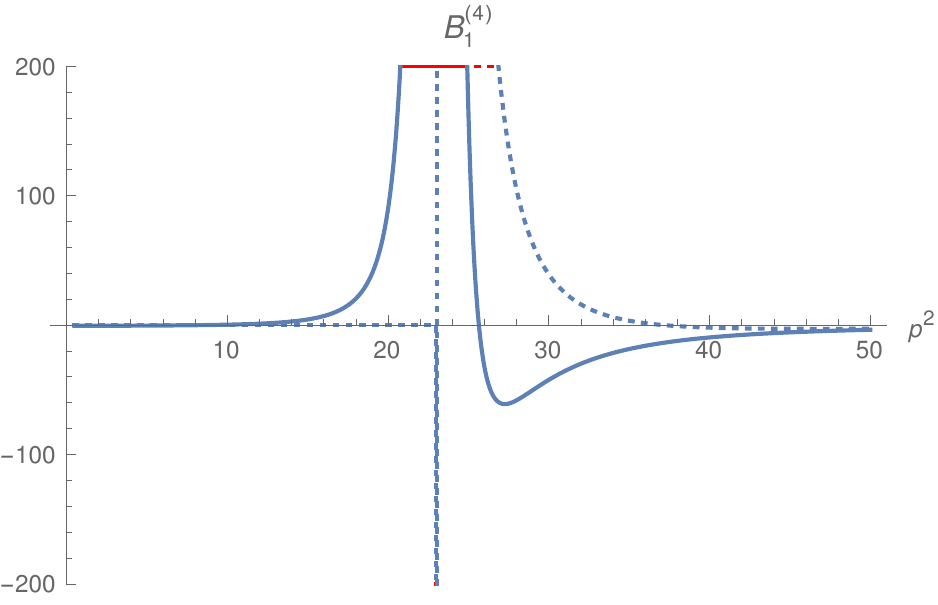} & \imgPlot{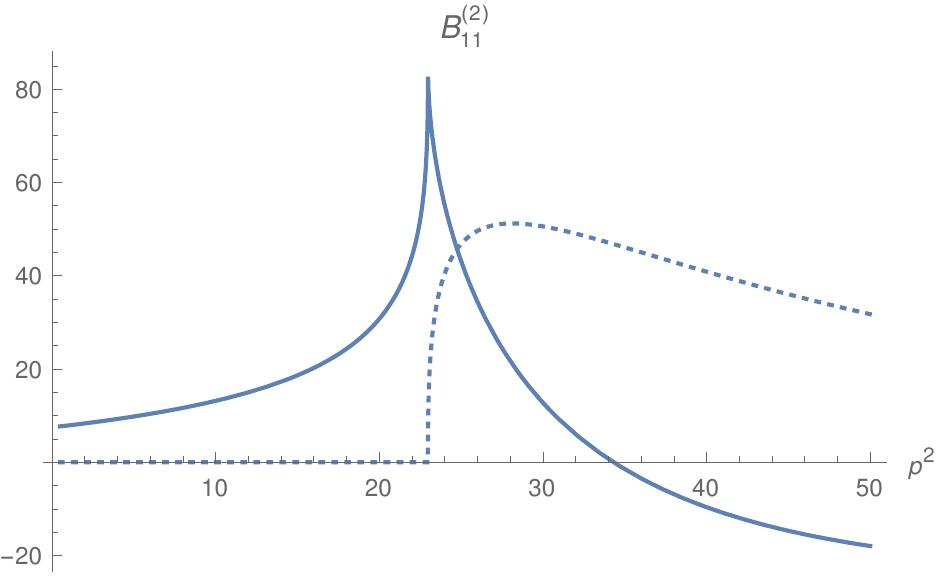} \\
        \imgPlot{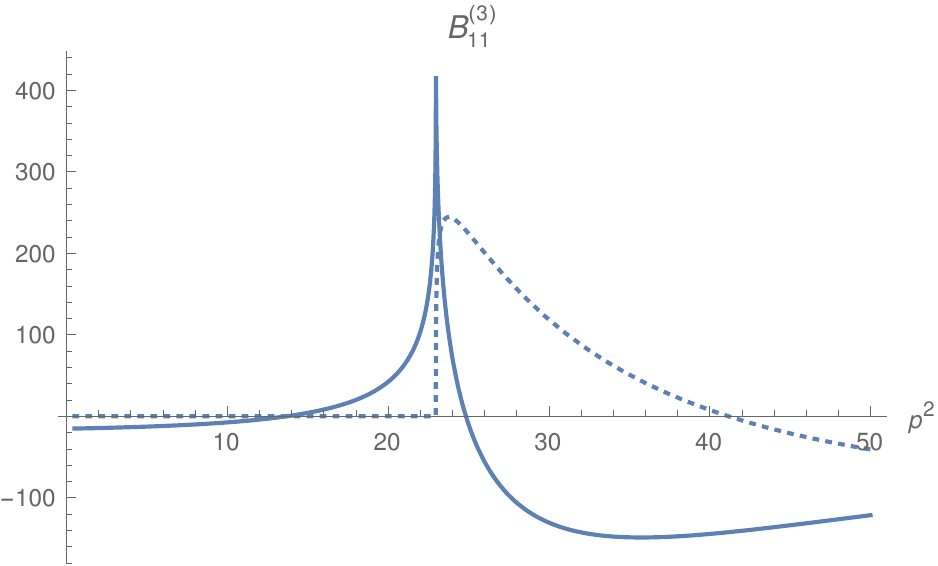} & \imgPlot{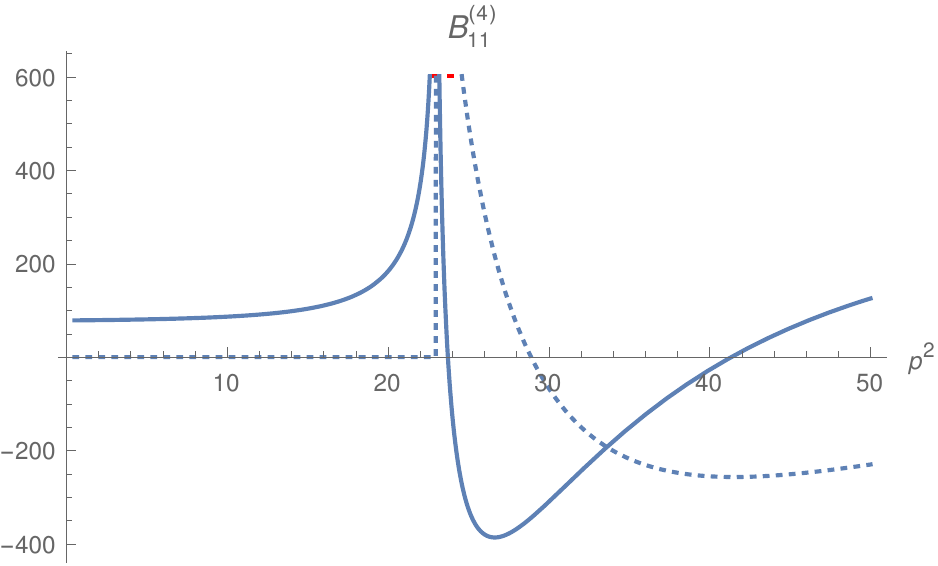}
 \end{tabular}
 \caption{Plots of some of the precanonical basis integrals of the unequal-mass three-loop banana family. Note that $B_1 = \epsilon(1+3\epsilon)I_{1122}^\text{banana}$ and that $B_{11} = \epsilon(1+3\epsilon)(1+4\epsilon)I_{1111}^\text{banana}$.}
 \label{fig:plots}
\end{figure}

In the following, we will denote the phase-space coordinates by $(p^2,m_1,m_2,m_3,m_4)$. As an illustrative example, we have computed results along the line:
\begin{align}
    \label{eq:uneqmassbananagamma1}
    \gamma(x) = (x, 2, 3/2, 4/3, 1)\,,
\end{align}
from $x = 1/2$ to $x = 50$. In Fig. \ref{fig:plots}, we provide plots for $B_1^{(2)}, B_1^{(3)}, B_1^{(4)}, B_{11}^{(2)}, B_{11}^{(3)}$ and $B_{11}^{(4)}$ along this line. 
These results were obtained in the following manner. First, we used the differential equations of the equal mass family to obtain high precision results at the point $(1/2,1,1,1,1)$. Next, we transported the results to the point $(1/2,2,3/2,4/3,1)$. Lastly we performed the expansions along the line $\gamma(x)$ to reach the point $(50,2,3/2,4/3,1)$. The transportation of the results along $\gamma(x)$ took about 1 hour and 5 minutes on a PC equipped with an i7-4700MQ processor. The expansions for the unequal-mass family were configured with the following options: 
\begin{minted}[linenos=false]{text}
    {
        ChopPrecision -> 250, DivisionOrder -> 4, EpsilonOrder -> 4,
        ExpansionOrder -> 70, RadiusOfConvergence -> 10, UseMobius -> True,
        UsePade -> True, WorkingPrecision -> 1000
    }
\end{minted}
The error reported by DiffExp at the point $(50,2,3/2,4/3,1)$ was of order $10^{-22}$. We performed an internal cross-check of the results by reaching the point $(50,2,3/2,4/3,1)$ through a different contour. In particular, we first used the differential equations of the equal-mass family to obtain results at high precision at the point $(50,1,1,1,1)$, and then we transported those to the point $(50,2,3/2,4/3,1)$ using the unequal-mass differential equations. We found that the maximum difference between the results at $(50,2,3/2,4/3,1)$ obtained along the different contours was of the order $10^{-24}$.

We also performed a higher precision evaluation along the line $\gamma(x)$. In this case we configured DiffExp with the following options for the unequal-mass family:
\begin{minted}[linenos=false]{text}
    {
        ChopPrecision -> 600, DivisionOrder -> 4, EpsilonOrder -> 4,
        ExpansionOrder -> 110, RadiusOfConvergence -> 10, UseMobius -> True,
        UsePade -> True, WorkingPrecision -> 1400
    }
\end{minted}
It took a bit under four hours to obtain the results along $\gamma(x)$. The error reported by DiffExp was of order $10^{-58}$. Upon cross-checking the results along an independent contour, like before, we found a maximum difference of order $10^{-61}$. Note that after the expansions are computed, it is almost instantaneous to evaluate the integrals anywhere along the line between $x=1/2$ and $x=50$, since this simply amounts to plugging numbers into the Padé approximants. For example, evaluating orders 0 to 4 in $\epsilon$ of all basis integrals from the Padé approximants, in the point $\gamma(10)$, takes about half a second. As a numerical example, we provide 55 digits after the decimal point of the coefficients in the $\epsilon$ expansion of the integral $B_{11}$ in the point $(50,2,3/2,4/3,1)$:
\begin{align}
    B_{11}^{(0)} &= 0 \nonumber\\
    B_{11}^{(1)} &= 5.1972521136965043170129578538563652405618939122389078645 \nonumber\\&\quad + i~6.8755169535390207501370685645538902299559024551830956594 \nonumber\\
    B_{11}^{(2)} &= -17.9580108112094060899523361698928478948780687053899075733 \nonumber\\&\quad + i~31.7436703633693090908402932299011971913508950649494231047 \nonumber\\
    B_{11}^{(3)} &= -121.5101152068177565203392807541216084962880772908306370668 \nonumber\\&\quad - i~40.7690762360202766453775999917172226537428258529145754746 \nonumber\\
    B_{11}^{(4)} &= 125.6113388023605534745593764004798958232118632681257073923 \nonumber\\&\quad - i~229.9200257172388589952062757571215176834471783495112755027
\end{align}
Note that it is considerably faster to reach the point $(50,2,3/2,4/3,1)$ if we move from the infinite momentum limit to the point $(50,1,1,1,1)$, and from there to the point $(50,2,3/2,4/3,1)$, instead of moving along the line $\gamma(x)$. The total time to reach $(50,2,3/2,4/3,1)$ from the infinite momentum limit is then around 23 minutes, at an estimated precision of $10^{-70}$. If we repeat the computation at a lower expansion order, we manage to achieve an estimated precision of $10^{-34}$ in 6 minutes. We performed a cross-check of the results against pySecDec \cite{Borowka:2017idc} in a few points, for which we obtained full agreement every time within the errors reported by pySecDec.

\subsection{Other examples}
We have tested DiffExp on the planar two-loop five-point one-mass integral families of Ref. \cite{Abreu:2020jxa}, taking the differential equations and boundary conditions from the ancillary files of the paper. The paper provides high-precision boundary conditions at seven points in phase-space, accurate up to at least 128 digits. Among other checks, we transported the numerical results for family "zzz" from phase-space point one to phase-space point two at a precision of at least 128 digits, finding full agreement. The computation took about 2 hours and 15 minutes to complete. We also transported the results at a lower expansion order from phase-space point one to phase-space point six, which yielded a maximum error of order $10^{-23}$, and which took a bit under half an hour to complete. The integral families of Ref. \cite{Abreu:2020jxa} can be computed with the notebook \texttt{5pPlanar1Mass.nb} in the \texttt{Examples} folder.

Furthermore, we have tested DiffExp on the two-loop five-point non-planar massless integrals of Ref. \cite{Chicherin:2018old}, using the differential equations from the ancillary files of that paper. The ancillary files of the paper provide numerical results at two points in phase-space at a precision of at least 50 digits. We cross-checked these results by transporting the results from one point to the other using DiffExp, finding agreement of at least 50 digits. The transportation of the results took a bit under five minutes to complete. The integrals of Ref. \cite{Chicherin:2018old} can be computed with the notebook \texttt{5pNonPlanar.nb} in the \texttt{Examples} folder.

\section{Conclusions and outlook}
\label{sec:conclusions}
In this paper we have presented the DiffExp Mathematica package for solving families of Feynman integrals in terms of truncated one-dimensional series expansions, through their systems of differential equations. DiffExp is built on the integration strategy that was developed in Ref. \cite{Francesco:2019yqt}, and which was further studied and applied in Refs. \cite{Bonciani:2019jyb, Frellesvig:2019byn}. The strategy has also recently been applied in Ref. \cite{Abreu:2020jxa}. DiffExp is the first publicly available Mathematica package that implements these methods. Compared to those papers, we made a few novel improvements. 

In Section \ref{sec:intorder}, we described how to automatically derive an integration sequence from the differential equations. In Section \ref{sec:gensols}, an optimized integration strategy was discussed for solving coupled integrals. In Section \ref{sec:predivision}, we discussed a segmentation strategy that is slightly improved from the one of Ref. \cite{Frellesvig:2019byn}, with better matching of neighbouring line segments. In addition, we reviewed series acceleration methods in Section \ref{sec:precisionandnumerics}, in particular Padé approximants and suitably defined Möbius transformations, which were applied before in Ref. \cite{Frellesvig:2019byn} and described here in more detail. Lastly, we have provided in Section \ref{sec:examples} the first application of the series expansion strategy of Ref. \cite{Francesco:2019yqt} to integrals that are coupled at higher degrees than two, by considering the three-loop equal-mass and unequal-mass banana graph families, for which the top sectors are coupled at order 3 and order 11 respectively. 

For future work, it would be interesting to extend DiffExp to work with bases of integrals whose prefactors contains functions beyond the rational functions and square roots, such as elliptic integrals, which appear in the canonical basis of the equal and unequal-mass sunrise family \cite{Adams:2018yfj, Bogner:2019lfa}.

\section*{Acknowledgements}
I would like to thank F. Moriello for introducing me to series expansions methods, for developing the first private implementation of these methods in Ref. \cite{Francesco:2019yqt}, and for numerous useful discussions on the topic. I would like to thank G. Salvatori, and L. Maestri for many useful discussions during the preparation of Refs. \cite{Bonciani:2019jyb, Frellesvig:2019byn} and onwards. I would like to thank R. Britto for providing helpful comments to improve the manuscript. This work was funded by the European Research Council (ERC) under grant agreement No. 647356 (CutLoops).

\bibliographystyle{unsrt}
\bibliography{refs}

\begin{thebibliography}{10}

\bibitem{Francesco:2019yqt}
Francesco Moriello.
\newblock {Generalised power series expansions for the elliptic planar families
  of Higgs + jet production at two loops}.
\newblock {\em JHEP}, 01:150, 2020.

\bibitem{Laporta:2001dd}
S.~Laporta.
\newblock {High precision calculation of multiloop Feynman integrals by
  difference equations}.
\newblock {\em Int. J. Mod. Phys. A}, 15:5087--5159, 2000.

\bibitem{Lee:2013mka}
Roman~N. Lee.
\newblock {LiteRed 1.4: a powerful tool for reduction of multiloop integrals}.
\newblock {\em J. Phys. Conf. Ser.}, 523:012059, 2014.

\bibitem{Smirnov:2019qkx}
A.~V. Smirnov and F.~S. Chuharev.
\newblock {FIRE6: Feynman Integral REduction with Modular Arithmetic}.
\newblock {\em Computer Physics Communications}, 247:106877, 2020.

\bibitem{Klappert:2020nbg}
Jonas Klappert, Fabian Lange, Philipp Maierh\"ofer, and Johann Usovitsch.
\newblock {Integral Reduction with Kira 2.0 and Finite Field Methods}.
\newblock 8 2020.

\bibitem{Smirnov:2015mct}
Alexander~V. Smirnov.
\newblock {FIESTA4: Optimized Feynman integral calculations with GPU support}.
\newblock {\em Comput. Phys. Commun.}, 204:189--199, 2016.

\bibitem{Borowka:2017idc}
S.~Borowka, G.~Heinrich, S.~Jahn, S.~P. Jones, M.~Kerner, J.~Schlenk, and
  T.~Zirke.
\newblock {pySecDec: a toolbox for the numerical evaluation of multi-scale
  integrals}.
\newblock {\em Comput. Phys. Commun.}, 222:313--326, 2018.

\bibitem{Vanhove:2018mto}
Pierre Vanhove.
\newblock {Feynman integrals, toric geometry and mirror symmetry}.
\newblock In {\em {Proceedings, KMPB Conference: Elliptic Integrals, Elliptic
  Functions and Modular Forms in Quantum Field Theory: Zeuthen, Germany,
  October 23-26, 2017}}, pages 415--458, 2019.

\bibitem{delaCruz:2019skx}
Leonardo de~la Cruz.
\newblock {Feynman integrals as A-hypergeometric functions}.
\newblock {\em JHEP}, 12:123, 2019.

\bibitem{Klausen:2019hrg}
René~Pascal Klausen.
\newblock {Hypergeometric Series Representations of Feynman Integrals by GKZ
  Hypergeometric Systems}.
\newblock {\em JHEP}, 04:121, 2020.

\bibitem{Feng:2019bdx}
Tai-Fu Feng, Chao-Hsi Chang, Jian-Bin Chen, and Hai-Bin Zhang.
\newblock {GKZ-hypergeometric systems for Feynman integrals}.
\newblock {\em Nucl. Phys.}, B953:114952, 2020.

\bibitem{Klemm:2019dbm}
Albrecht Klemm, Christoph Nega, and Reza Safari.
\newblock {The $l$-loop Banana Amplitude from GKZ Systems and relative
  Calabi-Yau Periods}.
\newblock {\em JHEP}, 04:088, 2020.

\bibitem{Bonisch:2020qmm}
Kilian B\"onisch, Fabian Fischbach, Albrecht Klemm, Christoph Nega, and Reza
  Safari.
\newblock {Analytic Structure of all Loop Banana Amplitudes}.
\newblock 8 2020.

\bibitem{Brown:2009ta}
Francis C.~S. Brown.
\newblock {On the periods of some Feynman integrals}.
\newblock 2009.

\bibitem{Panzer:2014caa}
Erik Panzer.
\newblock {Algorithms for the symbolic integration of hyperlogarithms with
  applications to Feynman integrals}.
\newblock {\em Comput. Phys. Commun.}, 188:148--166, 2015.

\bibitem{Primo:2016ebd}
Amedeo Primo and Lorenzo Tancredi.
\newblock {On the maximal cut of Feynman integrals and the solution of their
  differential equations}.
\newblock {\em Nucl. Phys.}, B916:94--116, 2017.

\bibitem{Goncharov:1998kja}
Alexander~B. Goncharov.
\newblock {Multiple polylogarithms, cyclotomy and modular complexes}.
\newblock {\em Math. Res. Lett.}, 5:497--516, 1998.

\bibitem{brown2011multiple}
Francis C.~S. Brown and Andrey Levin.
\newblock Multiple elliptic polylogarithms, 2011.

\bibitem{Adams:2014vja}
Luise Adams, Christian Bogner, and Stefan Weinzierl.
\newblock {The two-loop sunrise graph in two space-time dimensions with
  arbitrary masses in terms of elliptic dilogarithms}.
\newblock {\em J. Math. Phys.}, 55(10):102301, 2014.

\bibitem{Broedel:2014vla}
Johannes Broedel, Carlos~R. Mafra, Nils Matthes, and Oliver Schlotterer.
\newblock {Elliptic multiple zeta values and one-loop superstring amplitudes}.
\newblock {\em JHEP}, 07:112, 2015.

\bibitem{Broedel:2017kkb}
Johannes Broedel, Claude Duhr, Falko Dulat, and Lorenzo Tancredi.
\newblock {Elliptic polylogarithms and iterated integrals on elliptic curves.
  Part I: general formalism}.
\newblock {\em JHEP}, 05:093, 2018.

\bibitem{Broedel:2017siw}
Johannes Broedel, Claude Duhr, Falko Dulat, and Lorenzo Tancredi.
\newblock {Elliptic polylogarithms and iterated integrals on elliptic curves
  II: an application to the sunrise integral}.
\newblock {\em Phys. Rev.}, D97(11):116009, 2018.

\bibitem{Adams:2018yfj}
Luise Adams and Stefan Weinzierl.
\newblock {The $\varepsilon$-form of the differential equations for Feynman
  integrals in the elliptic case}.
\newblock {\em Phys. Lett.}, B781:270--278, 2018.

\bibitem{Bogner:2019lfa}
Christian Bogner, Stefan Müller-Stach, and Stefan Weinzierl.
\newblock {The unequal mass sunrise integral expressed through iterated
  integrals on $\overline{\mathcal M}_{1,3}$}.
\newblock {\em Nucl. Phys.}, B954:114991, 2020.

\bibitem{Zhang:2016hdb}
Yang Zhang and Alessandro Georgoudis.
\newblock {Integral reduction via algebraic curves}.
\newblock {\em PoS}, RADCOR2015:085, 2016.

\bibitem{Besier:2018jen}
Marco Besier, Duco Van~Straten, and Stefan Weinzierl.
\newblock {Rationalizing roots: an algorithmic approach}.
\newblock {\em Commun. Num. Theor. Phys.}, 13:253--297, 2019.

\bibitem{Besier:2019kco}
Marco Besier, Pascal Wasser, and Stefan Weinzierl.
\newblock {RationalizeRoots: Software Package for the Rationalization of Square
  Roots}.
\newblock {\em Comput. Phys. Commun.}, 253:107197, 2020.

\bibitem{Heller:2019gkq}
Matthias Heller, Andreas von Manteuffel, and Robert~M. Schabinger.
\newblock {Multiple polylogarithms with algebraic arguments and the two-loop
  EW-QCD Drell-Yan master integrals}.
\newblock 2019.

\bibitem{Besier:2020hjf}
Marco Besier and Dino Festi.
\newblock {Rationalizability of square roots}.
\newblock 6 2020.

\bibitem{Brown:2020rda}
Francis Brown and Claude Duhr.
\newblock {A double integral of dlog forms which is not polylogarithmic}.
\newblock 6 2020.

\bibitem{Bonciani:2016qxi}
Roberto Bonciani, Vittorio Del~Duca, Hjalte Frellesvig, Johannes~M. Henn,
  Francesco Moriello, and Vladimir~A. Smirnov.
\newblock {Two-loop planar master integrals for Higgs$\to 3$ partons with full
  heavy-quark mass dependence}.
\newblock {\em JHEP}, 12:096, 2016.

\bibitem{Bonciani:2019jyb}
R.~Bonciani, V.~Del~Duca, H.~Frellesvig, J.~M. Henn, M.~Hidding, L.~Maestri,
  F.~Moriello, G.~Salvatori, and V.~A. Smirnov.
\newblock {Evaluating a family of two-loop non-planar master integrals for
  Higgs + jet production with full heavy-quark mass dependence}.
\newblock {\em JHEP}, 01:132, 2020.

\bibitem{Frellesvig:2019byn}
Hjalte Frellesvig, Martijn Hidding, Leila Maestri, Francesco Moriello, and
  Giulio Salvatori.
\newblock {The complete set of two-loop master integrals for Higgs + jet
  production in QCD}.
\newblock 2019.

\bibitem{Abreu:2020jxa}
Samuel Abreu, Harald Ita, Francesco Moriello, Ben Page, Wladimir Tschernow, and
  Mao Zeng.
\newblock {Two-Loop Integrals for Planar Five-Point One-Mass Processes}.
\newblock 2020.

\bibitem{chengwu}
Hung Cheng and Tai~Tsun Wu.
\newblock {\em Expanding Protons: Scattering at High Energies}.
\newblock MIT Press, Cambridge, MA, USA, 1987.

\bibitem{Bogner:2010kv}
Christian Bogner and Stefan Weinzierl.
\newblock {Feynman graph polynomials}.
\newblock {\em Int. J. Mod. Phys.}, A25:2585--2618, 2010.

\bibitem{Henn:2013nsa}
Johannes~M. Henn, Alexander~V. Smirnov, and Vladimir~A. Smirnov.
\newblock {Evaluating single-scale and/or non-planar diagrams by differential
  equations}.
\newblock {\em JHEP}, 03:088, 2014.

\bibitem{Panzer:2014gra}
Erik Panzer.
\newblock {On hyperlogarithms and Feynman integrals with divergences and many
  scales}.
\newblock {\em JHEP}, 03:071, 2014.

\bibitem{Binoth:2000ps}
T.~Binoth and G.~Heinrich.
\newblock {An automatized algorithm to compute infrared divergent multiloop
  integrals}.
\newblock {\em Nucl. Phys.}, B585:741--759, 2000.

\bibitem{Binoth:2003ak}
T.~Binoth and G.~Heinrich.
\newblock {Numerical evaluation of multiloop integrals by sector
  decomposition}.
\newblock {\em Nucl. Phys.}, B680:375--388, 2004.

\bibitem{Bogner:2007cr}
Christian Bogner and Stefan Weinzierl.
\newblock {Resolution of singularities for multi-loop integrals}.
\newblock {\em Comput. Phys. Commun.}, 178:596--610, 2008.

\bibitem{Landau:1959fi}
L.D. Landau.
\newblock {On analytic properties of vertex parts in quantum field theory}.
\newblock {\em Nucl. Phys.}, 13(1):181--192, 1960.

\bibitem{Beneke:1997zp}
M.~Beneke and Vladimir~A. Smirnov.
\newblock {Asymptotic expansion of Feynman integrals near threshold}.
\newblock {\em Nucl. Phys.}, B522:321--344, 1998.

\bibitem{Smirnov:1999bza}
Vladimir~A. Smirnov.
\newblock {Problems of the strategy of regions}.
\newblock {\em Phys. Lett.}, B465:226--234, 1999.

\bibitem{Jantzen:2011nz}
Bernd Jantzen.
\newblock {Foundation and generalization of the expansion by regions}.
\newblock {\em JHEP}, 12:076, 2011.

\bibitem{Semenova:2018cwy}
Tatiana~Yu Semenova, Alexander~V. Smirnov, and Vladimir~A. Smirnov.
\newblock {On the status of expansion by regions}.
\newblock {\em Eur. Phys. J.}, C79(2):136, 2019.

\bibitem{Ananthanarayan:2018tog}
B.~Ananthanarayan, Abhishek Pal, S.~Ramanan, and Ratan Sarkar.
\newblock {Unveiling Regions in multi-scale Feynman Integrals using
  Singularities and Power Geometry}.
\newblock {\em Eur. Phys. J. C}, 79(1):57, 2019.

\bibitem{Ananthanarayan:2020ptw}
B.~Ananthanarayan, Abhijit~B. Das, and Ratan Sarkar.
\newblock {Asymptotic Analysis of Feynman Diagrams and their Maximal Cuts}.
\newblock 3 2020.

\bibitem{Pak:2010pt}
A.~Pak and A.~Smirnov.
\newblock {Geometric approach to asymptotic expansion of Feynman integrals}.
\newblock {\em Eur. Phys. J.}, C71:1626, 2011.

\bibitem{Jantzen:2012mw}
Bernd Jantzen, Alexander~V. Smirnov, and Vladimir~A. Smirnov.
\newblock {Expansion by regions: revealing potential and Glauber regions
  automatically}.
\newblock {\em Eur. Phys. J.}, C72:2139, 2012.

\bibitem{Kotikov:1990kg}
A.~V. Kotikov.
\newblock {Differential equations method: New technique for massive Feynman
  diagrams calculation}.
\newblock {\em Phys. Lett.}, B254:158--164, 1991.

\bibitem{Kotikov:1991pm}
A.~V. Kotikov.
\newblock {Differential equation method: The Calculation of N point Feynman
  diagrams}.
\newblock {\em Phys. Lett.}, B267:123--127, 1991.
\newblock [Erratum: Phys. Lett.B295,409(1992)].

\bibitem{Kotikov:1991hm}
A.~V. Kotikov.
\newblock {Differential equations method: The Calculation of vertex type
  Feynman diagrams}.
\newblock {\em Phys. Lett.}, B259:314--322, 1991.

\bibitem{Henn:2013pwa}
Johannes~M. Henn.
\newblock {Multiloop integrals in dimensional regularization made simple}.
\newblock {\em Phys. Rev. Lett.}, 110:251601, 2013.

\bibitem{Pozzorini:2005ff}
S.~Pozzorini and E.~Remiddi.
\newblock {Precise numerical evaluation of the two loop sunrise graph master
  integrals in the equal mass case}.
\newblock {\em Comput. Phys. Commun.}, 175:381--387, 2006.

\bibitem{Aglietti:2007as}
U.~Aglietti, R.~Bonciani, L.~Grassi, and E.~Remiddi.
\newblock {The Two loop crossed ladder vertex diagram with two massive
  exchanges}.
\newblock {\em Nucl. Phys.}, B789:45--83, 2008.

\bibitem{Mueller:2015lrx}
Romain Mueller and Deniz~Gizem Öztürk.
\newblock {On the computation of finite bottom-quark mass effects in Higgs
  boson production}.
\newblock {\em JHEP}, 08:055, 2016.

\bibitem{Melnikov:2016qoc}
Kirill Melnikov, Lorenzo Tancredi, and Christopher Wever.
\newblock {Two-loop $gg \to Hg$ amplitude mediated by a nearly massless quark}.
\newblock {\em JHEP}, 11:104, 2016.

\bibitem{Lee:2017qql}
Roman~N. Lee, Alexander~V. Smirnov, and Vladimir~A. Smirnov.
\newblock {Solving differential equations for Feynman integrals by expansions
  near singular points}.
\newblock {\em JHEP}, 03:008, 2018.

\bibitem{Melnikov:2017pgf}
Kirill Melnikov, Lorenzo Tancredi, and Christopher Wever.
\newblock {Two-loop amplitudes for $q g \to H q$ and $q \bar{q} \to H g$
  mediated by a nearly massless quark}.
\newblock {\em Phys. Rev.}, D95(5):054012, 2017.

\bibitem{Lee:2018ojn}
Roman~N. Lee, Alexander~V. Smirnov, and Vladimir~A. Smirnov.
\newblock {Evaluating ‘elliptic’ master integrals at special kinematic
  values: using differential equations and their solutions via expansions near
  singular points}.
\newblock {\em JHEP}, 07:102, 2018.

\bibitem{Bonciani:2018uvv}
Roberto Bonciani, Giuseppe Degrassi, Pier~Paolo Giardino, and Ramona Gröber.
\newblock {A Numerical Routine for the Crossed Vertex Diagram with a
  Massive-Particle Loop}.
\newblock {\em Comput. Phys. Commun.}, 241:122--131, 2019.

\bibitem{Mistlberger:2018etf}
Bernhard Mistlberger.
\newblock {Higgs boson production at hadron colliders at N$^{3}$LO in QCD}.
\newblock {\em JHEP}, 05:028, 2018.

\bibitem{Bonciani:2018omm}
Roberto Bonciani, Giuseppe Degrassi, Pier~Paolo Giardino, and Ramona Gröber.
\newblock {Analytical Method for Next-to-Leading-Order QCD Corrections to
  Double-Higgs Production}.
\newblock {\em Phys. Rev. Lett.}, 121(16):162003, 2018.

\bibitem{Bruser:2018jnc}
Robin Brüser, Simon Caron-Huot, and Johannes~M. Henn.
\newblock {Subleading Regge limit from a soft anomalous dimension}.
\newblock {\em JHEP}, 04:047, 2018.

\bibitem{Davies:2018ood}
Joshua Davies, Go~Mishima, Matthias Steinhauser, and David Wellmann.
\newblock {Double-Higgs boson production in the high-energy limit: planar
  master integrals}.
\newblock {\em JHEP}, 03:048, 2018.

\bibitem{Davies:2018qvx}
Joshua Davies, Go~Mishima, Matthias Steinhauser, and David Wellmann.
\newblock {Double Higgs boson production at NLO in the high-energy limit:
  complete analytic results}.
\newblock {\em JHEP}, 01:176, 2019.

\bibitem{Coddington}
E.~Coddington and N.~Levinson.
\newblock {\em {Theory of ordinary differential equations}}.
\newblock 1955.

\bibitem{bender2013advanced}
Carl~M Bender and Steven~A Orszag.
\newblock {\em Advanced mathematical methods for scientists and engineers I:
  Asymptotic methods and perturbation theory}.
\newblock Springer Science \& Business Media, 2013.

\bibitem{Maierhoefer:2017hyi}
Philipp Maierh\"ofer, Johann Usovitsch, and Peter Uwer.
\newblock {Kira\textemdash{}A Feynman integral reduction program}.
\newblock {\em Comput. Phys. Commun.}, 230:99--112, 2018.

\bibitem{Maierhofer:2018gpa}
Philipp Maierhöfer and Johann Usovitsch.
\newblock {Kira 1.2 Release Notes}.
\newblock 2018.

\bibitem{Chicherin:2018old}
D.~Chicherin, T.~Gehrmann, J.~M. Henn, P.~Wasser, Y.~Zhang, and S.~Zoia.
\newblock {All Master Integrals for Three-Jet Production at
  Next-to-Next-to-Leading Order}.
\newblock {\em Phys. Rev. Lett.}, 123(4):041603, 2019.

\end{thebibliography}

\end{document}